\documentclass[%
reprint,
superscriptaddress,
nofootinbib,
nobibnotes,
aps,
prb,
floatfix,
]{revtex4-2}

\pdfoutput=1

\usepackage{amsmath}
\usepackage{amssymb}
\usepackage{graphicx}
\usepackage{dcolumn}
\usepackage{bbm}
\usepackage{bbold}
\usepackage{bm}
\usepackage[dvipsnames,x11names]{xcolor}
\usepackage[colorlinks=true,
	        linkcolor=hblue,
	        citecolor=hgreen,
	        filecolor=hblue,
	        urlcolor=hred
	        ]{hyperref}
\usepackage{mathtools}
\usepackage{booktabs}
\usepackage{mathrsfs}
\usepackage{orcidlink}
\usepackage{slashed}
\usepackage[protrusion=true,expansion,kerning=true,tracking=true,final]{microtype}
\usepackage{soul}
\usepackage{tabularx}
\usepackage{comment}

\definecolor{hgreen}{rgb}{0,.3,0}
\definecolor{hred}{rgb}{.3,0,0}
\definecolor{orange}{rgb}{1,0.5,0}
\definecolor{hblue}{rgb}{0,0,.3}
\definecolor{LightGray}{gray}{0.95}
\definecolor{gray}{gray}{0.6}

\DeclareOldFontCommand{\rm}{\normalfont\rmfamily}{\mathrm}
\DeclareOldFontCommand{\sf}{\normalfont\sffamily}{\mathsf}
\DeclareOldFontCommand{\tt}{\normalfont\ttfamily}{\mathtt}
\DeclareOldFontCommand{\bf}{\normalfont\bfseries}{\mathbf}
\DeclareOldFontCommand{\it}{\normalfont\itshape}{\mathit}
\DeclareOldFontCommand{\sl}{\normalfont\slshape}{\@nomath\sl}
\DeclareOldFontCommand{\sc}{\normalfont\scshape}{\@nomath\sc}

\newcommand{\Lagr}{\mathscr{L}}
\newcommand{\Lag}{\mathscr{L}}
\newcommand{\SO}{\mathrm{SO}}
\newcommand{\U}{\mathrm{U}}
\newcommand{\grO}{\mathrm{O}}

\newcommand{\eq}[1]{\eqref{eq:#1}}

\begin{document}

\title{Quantum multicriticality and emergent symmetry in Dirac systems\\with two order parameters at three-loop order}
\date{\today}

\author{Max~Uetrecht\,\orcidlink{0000-0001-8685-2543}}
\email{max.uetrecht@tu-dortmund.de}
\affiliation{Fakult\"at Physik, Technische Universit\"at Dortmund, D-44221 Dortmund, Germany}

\author{Igor F.~Herbut\,\orcidlink{0000-0001-5496-8330}}
\email{igor\_herbut@sfu.ca}
\affiliation{Department of Physics, Simon Fraser University, Burnaby, British Columbia, Canada V5A 1S6}
\affiliation{Institute for Solid State Physics, University of Tokyo, Kashiwa, 277-8581, Japan}
\affiliation{Institut f\"ur Theoretische Physik und Astrophysik, Universit\"at W\"urzburg, D-97074 W\"urzburg, Germany}
\affiliation{W\"urzburg-Dresden Cluster of Excellence ct.qmat, Am Hubland, D-97074 W\"urzburg, Germany}

\author{Michael~M.~Scherer\,\orcidlink{0000-0003-0766-9949}}
\email{scherer@tp3.rub.de}
\affiliation{Theoretische Physik III, Ruhr-Universit\"at Bochum, D-44801 Bochum, Germany}

\author{Emmanuel Stamou\,\orcidlink{0000-0002-8385-6159}}
\email{emmanuel.stamou@tu-dortmund.de}
\affiliation{Fakult\"at Physik, Technische Universit\"at Dortmund, D-44221 Dortmund, Germany}

\author{Tom~Steudtner\,\orcidlink{0000-0003-1935-0417}}
\email{tom2.steudtner@tu-dortmund.de}
\affiliation{Fakult\"at Physik, Technische Universit\"at Dortmund, D-44221 Dortmund, Germany}

\begin{abstract}
Two-dimensional materials with interacting Dirac excitations can host quantum
multicritical behavior near the phase boundaries of the semimetallic and 
two-ordered phases.  
We study such behavior in Gross--Neveu--Yukawa field theories where
$N_f$~flavors of Dirac fermions are coupled to two order-parameter fields with
$\SO(N_A)$ and $\SO(N_B)$ symmetry, respectively.  
To that end, we employ the perturbative renormalization group up to three-loop
order in $4-\epsilon$ spacetime dimensions.  We distinguish two key scenarios:
{(i)}~The two orders are compatible as characterized by anticommuting mass terms,
and {(ii)}~the orders are incompatible.  
For the first case, we explore the stability of a quantum multicritical point
with emergent $\SO(N_A\!+\!N_B)$~symmetry.  
We find that the stability is controlled by increasing the number of Dirac
fermion flavors.
Moreover, we extract the series expansion of the leading critical exponents for
the chiral $\SO(4)$ and $\SO(5)$~models up to third order in~$\epsilon$.  
Notably, we find a tendency towards rapidly growing expansion coefficients
at higher orders, rendering an extrapolation to $\epsilon=1$ difficult.  
For the second scenario, we study a model with $\SO(4) \simeq \SO(3) \times \SO(3)$
symmetry, which was recently suggested to describe criticality of
antiferromagnetism and superconductivity in Dirac  systems.  
However, it was also argued that a physically admissible renormalization-group
fixed point only exists for $N_f$ above a critical number $N_{c}^>$.  
We determine the corresponding series expansion at three-loop order as
$N_{c}^>\approx 16.83-7.14\epsilon-7.12\epsilon^2$.  
This suggests that the physical choice of $N_f=2$ may be a borderline case,
where true criticality and pseudocriticality, as induced by fixed-point
annihilation, are extremely challenging to distinguish.
\end{abstract}

\maketitle

\section{Introduction\label{sec:introduction}}

The understanding of systems with many degrees of freedom, e.g., quantum
materials, statistical models, or quantum field theories, fundamentally relies
on the identification of their underlying symmetries.
Theoretical descriptions typically take such symmetries into account by
incorporating the relevant ones in the formulation of the system's Hamiltonian
that defines the microscopic starting point of a theory.
On a macroscopic scale, symmetries can be spontaneously broken through the
formation of different types of order, which can even compete or show a complex
interplay with each other.
Interestingly, in the presence of various ordering tendencies it can also
happen that the microscopic description has lower symmetry than the macroscopic
one, i.e., there are cases that exhibit an {\itshape emergent} higher symmetry.  
For example, in the context of high-temperature superconductors, it was
suggested that aspects of their complex phase structure can be understood from
an enhanced $\SO(5)$~symmetry, emerging from the combination of the
$\SO(3)$~antiferromagnetic and $\mathrm{U}(1)$~superconducting order~\cite{sczhang1997}.  
Similarly, there is evidence of an emergent $\SO(5)$~symmetry in numerical
simulations of deconfined quantum critical
points~\cite{PhysRevLett.115.267203,PhysRevB.99.195110,PhysRevX.7.031052},
which can be further substantiated by duality conjectures~\cite{PhysRevX.7.031051}.

Despite the evidence for the possibility of emergent symmetry in systems with
two coupled orders, simple statistical models, which only take into account
order-parameter fluctuations, generally do not exhibit this
phenomenon~\cite{PELISSETTO2002549,herbut2007modern,PhysRevE.88.042141}.
For example, a system of two coupled order-parameter fields with $\SO(N_A)$ and
$\SO(N_B)$~symmetry, respectively, does not appear to feature a natural emergent
$\SO(N_A+N_B)$~symmetry for $N_A+N_B\geq 3$~\cite{PhysRevB.67.054505}.

Interestingly, recent studies collected evidence that gapless Dirac fermions as
they appear in various materials~\cite{wehling2014dirac,vafek2014dirac} support
the formation of different kinds of emergent
symmetries~\cite{Ponte:2012ru,doi:10.1126/science.1248253,PhysRevLett.114.237001,Roy:2015zna,PhysRevB.94.205136,PhysRevB.96.115132,li2017fermion,Yao2017,jian2017fermion,PhysRevB.96.195162,PhysRevB.108.L161108,PhysRevD.109.096026,Han:2024swe,PhysRevB.97.125137,PhysRevB.84.113404,PhysRevB.97.041117,PhysRevB.97.205117,PhysRevResearch.2.022005}.
Specifically, for the case of $N_f$ flavors of Dirac fermions coupled to
several order-parameter
fields~\cite{Boyack:2020xpe,Herbut:2023xgz,PhysRevB.84.113404,PhysRevB.92.035429,PhysRevB.93.125119,PhysRevB.97.041117,PhysRevB.97.205117,PhysRevB.99.241103,PhysRevResearch.2.022005,Herbut:2022zzw,Uetrecht:2023uou,Fornoville:2025rdd}
with $\SO(N_A)$ and $\SO(N_B)$~symmetry, respectively, it has been argued that a
set of two compatible orders, characterized by anticommuting mass terms of the
Dirac excitations, gives rise to a stable quantum multicritical point with
emergent $\SO(N_A+N_B)$~symmetry~\cite{PhysRevB.97.041117,PhysRevB.97.205117}.

So far, however, the emergence of $\SO(N_A\!+\!N_B)$~symmetry at such a quantum
multicritical point has only been studied at one-loop order with the
perturbative renormalization group and higher-loop corrections might be
detrimental to the stability of the emergent symmetry, as known from
statistical models~\cite{PELISSETTO2002549,PhysRevB.67.054505}

Here, we expand upon that previous work by providing a complete analysis of an
extended model of $N_f$ flavors of Dirac fermions, Yukawa-coupled to two
order-parameters at up to three loops in $4-\epsilon$ dimensions.
Within the framework of the $\epsilon$ expansion, we extrapolate our results to
the physical case of $2+1$ spacetime dimensions, which are relevant for the
relativistic Mott
transitions~\cite{PhysRevLett.97.146401,PhysRevB.80.075432,Boyack:2020xpe}
appearing in highly tunable Dirac systems engineered from moir\'e
semiconductors~\cite{ma2024relativistic,yang2025correlated,Hawashin:2025cua,Tolosa-Simeon:2025fot}.

\subsection{Outline}

We start our study in Sec.~\ref{sec:modeld3} by reviewing and expanding upon
the previously explored class of Gross--Neveu--Yukawa models. We first discuss 
the models with emergent $\SO(N_A\!+\!N_B)$ symmetry and subsequently the 
less symmetric $\SO(N_A) \times \SO(N_B)$ cases.
After establishing conditions for the emergence of the larger $\SO(N_A\!+\!N_B)$
symmetry, we discuss the analytic continuation of the model to $d=4-\epsilon$
in Sec.~\ref{sec:model4d}.
In Sec.~\ref{sec:analysis} and Sec.~\ref{subsec:compatible} we present its
$\beta$ functions at three-loop order, which are computed here for the first time, 
and discuss the stability of the fixed point compatible with the emergent 
symmetry for general values of~$N_A$, $N_B$, and~$N_f$.
As a corollary we also provide in Sec.~\ref{sec:incompatible} higher-order
results for a related model with $\SO(4) \simeq \SO(3) \times \SO(3)$ symmetry,
which was recently suggested to describe criticality of antiferromagnetism and
superconductivity in Dirac
systems~\cite{PhysRevLett.128.117202,Herbut:2022zzw,Uetrecht:2023uou}. We
conclude in Sec.~\ref{sec:conclusion}.

\section{Gross--Neveu--Yukawa with two order parameters in \texorpdfstring{$\boldsymbol{d=3}$}{d=3}\label{sec:modeld3}}

In this section, we review and expand upon the essentials of the
Gross--Neveu--Yukawa (GNY)
field theories with two compatible order-parameter fields studied in
Refs.~\cite{PhysRevB.97.041117,PhysRevB.97.205117,PhysRevResearch.2.022005}.
Therein, GNY models have been employed to describe the interaction of
spin-$1/2$ fermions on the honeycomb lattice at charge neutrality and
in the vicinity of phase transitions towards two different types of order,
described by $\SO(N_A)$ and $\SO(N_B)$ symmetric order-parameter fields.
Our focus here is the effect that fermions can have in stabilizing a
larger, emergent $\SO(N_A\!+\!N_B)$ global symmetry at the quantum critical point.
The emergence of an $\SO(N_A\!+\!N_B)$ is only possible if the underlying
$\SO(N_A) \times \SO(N_B)$ symmetric model is compatible with the enlarged symmetry.
We shall see that this provides strong constraints on the fermionic field
content and its transformation properties. To best illustrate this we begin
with the theory in which the enlarged symmetry is manifest.

\subsection{\texorpdfstring{$\SO(N_A\!+\!N_B)$}{SO(NA + NB)} GNY Theory}

The starting point is a $(2+1)$-dimensional theory with a fermion field,
$\psi$, with $d_\gamma$ complex components that satisfy a $d_\gamma\times
d_\gamma$-dimensional Clifford algebra with $N_A\!+\!N_B\!+\!2$ elements, i.e., $\psi$
is a Dirac fermion and 
\begin{equation}
  \{\gamma^m,\gamma^n\} = 2 \delta^{mn} \mathbb{1}_{d_\gamma \times d_\gamma}\,,
  \label{eq:Clif}
\end{equation}
with $m,n=1,\dots,N_A\!+\!N_B\!+\!2$, and with the ``$2$'' elements being related to the invariance  of the free Hamiltonian under rotations in $d_s=2$ space
dimensions~\cite{PhysRevB.79.085116,PhysRevB.97.041117,Boyack:2020xpe,Herbut:2023xgz}.
To realize the full Clifford algebra we need a sufficient number 
of fermionic components \cite{PhysRevB.97.041117}, namely\footnote{
$\lfloor \cdot \rfloor$ denotes the floor function, $\lfloor x \rfloor =\max\{ n\in \mathbb{Z}~\text{for}~n\leq x\}$. 
}

\begin{equation}\label{eq:compatible}
  d_\gamma \geq 2^{\left\lfloor\frac{N_A+N_B+2}{2}\right\rfloor} = 2\times 2^{\left\lfloor\frac{N_A+N_B}{2}\right\rfloor}
\end{equation}
and with $d_\gamma$ a multiple of four, i.e.,  $d_\gamma= 4 N_f$ with $N_f\in\mathbb{N}^+$,
such that the definition of $N_f$ is  consistent with Ref.~\cite{PhysRevB.97.041117}.
This na\"ively appears to imply that $\psi$ transforms as a spinor of an
$\SO(N_A\!+\!N_B\!+2)$.\footnote{
Throughout this work we refer to the spinor representation of $\SO(N)$ as the 
projective representation that corresponds to the fundamental representation of 
$\text{Spin}(M)$, i.e., the double cover of $\SO(N)$. The spinor representation
is irreducible if $N$ is odd and reducible if $N$ is even. Its 
generators are given by $S^{m n} = \frac{i}{2} [\gamma^m, \gamma^n]$
with $n,m=1,\dots,N$ and with the $\gamma$s' satisfying the Clifford algebra
$\{\gamma^m, \gamma^n\} = 2 \delta^{mn} \mathbb{1}$.
The dimension of the spinor representation 
is $2^{\left\lfloor\frac{N}{2}\right\rfloor}$. For more details see, e.g., Ref.~\cite{Zee:2016fuk}.}
However, the number of required $d_\gamma$  components is the same also 
if $\psi$ transforms in the spinor representation of both a global $\SO(N_A\!+\!N_B)$ 
and an internal $\SO(3)$ symmetry, i.e., a symmetry that operates also on (Euclidean) spacetime.
This is a consequence of the floor function above and thus only true for even
space dimensions. As we shall see below, the implication thereof is that the
Clifford algebra is the same also in an $\SO(N_A\!+\!N_B)\times \SO(3)$--invariant
theory. The first part, enables the coupling of $\psi$ with the order
parameter, which transforms under the $\SO(N_A\!+\!N_B)$ but not under the 
internal $\SO(3)$ symmetry.

$N_A\!+\!N_B$ of the elements of the Clifford algebra in Eq.~\eq{Clif} are
associated to the $\SO(N_A\!+\!N_B)$ symmetry under which the order parameter
$\vec\Phi_{A+B}$ is taken to transform in the fundamental representation;
$\vec\Phi_{A+B}$ has $N_A\!+\!N_B$ real components. Restricting
Eq.~\eq{Clif} to these elements shows that they satisfy their own
$\SO(N_A\!+\!N_B)$ Clifford algebra. Similarly, also the remaining two
elements satisfy their own Clifford algebra.
In fact when the internal $\SO(3)$ is imposed, there is an additional,
{\itshape independent} $\gamma$ matrix fulfilling the Clifford algebra. 
Nevertheless, as seen from Eq.~\eqref{eq:compatible}, $\SO(3)$ invariance 
does not require altering the minimal number of fermionic degrees of freedom 
necessary for invariance under rotations in $d_s=2$ space
dimensions.

This leads us to consider an $\SO(N_A\!+\!N_B)$--invariant 
GNY Lagrangian in $2+1$ spacetime dimensions in which
the $\vec{\Phi}_{A+B}$ order parameter is coupled to 
the $d_\gamma$-component complex fermion fields $\psi$. 
In Euclidean signature it takes the form
\begin{align}
\Lagr^{A+B}_{2+1} 
  &= 
  \frac{1}{2} (\partial_\mu \vec{\Phi}_{A+B})^2 
  -\frac{r}{2} (\vec{\Phi}_{A+B})^2
    -\frac{\lambda}{8} (\vec{\Phi}_{A+B})^4\nonumber\\
  &+
    \psi^{\dagger} \partial_\tau \psi 
    + \psi^{\dagger}\Big[\mathbb{1}_{\frac{d_\gamma}{2}\!\times\!\frac{d_\gamma}{2}} \otimes \sum_{k=1,2} \sigma_k \left(-i \partial _k\right)\Big] \psi \label{eq:Lag3dAB}\\
  & - \psi^{\dagger} \big[ g\,\vec{\Phi}_{A+B}\!\cdot\!\vec\beta_{A+B}\big]   \psi \,,\nonumber
\end{align}
with $\sigma_k$ the $2 \times 2$ Pauli matrices satisfying the Clifford 
algebra $\{\sigma_i, \sigma_j\} = 2 \delta_{ij} \mathbb{1}_{2\times 2}$.
The first and second lines are directly invariant under the internal $\SO(3)$
symmetry, i.e., they are Lorentz invariant (for details see Appendix A in
Ref.~\cite{Uetrecht:2023uou}).
The model thus possesses the effective version of Lorentz symmetry that has 
been argued to emerge in two-dimensional Dirac systems 
at criticality~\cite{Roy:2015zna}.
Note that this Lagrangian has only one Yukawa coupling~$g$. For cases at which $N_f$ is larger than its minimal value, cf. Eq.~\eqref{eq:compatible}, this can always be imposed by flavor symmetry as explicitly constructed in Sec.~\ref{sec:model4d}.

The $d_\gamma \times d_\gamma$ matrices, $\vec{\beta}_{A+B}$, are the $N_A\!+\!N_B$ 
matrices in $\vec\gamma$ that satisfy the 
$\SO(N_A\!+\!N_B)$ Clifford algebra.
They facilitate the only $\SO(N_A\!+\!N_B)$--invariant coupling of the fermions to the 
order parameter $\vec{\Phi}_{A+B}$.\footnote{
\label{fn:gammaF}
If $N_A\! +\! N_B$ is even, the Clifford algebra admits an
additional Gamma matrix, $\gamma_F$, that anticommutes with all
$\vec\gamma_{A+B}$, i.e., $\{\gamma_F, \gamma_{A+B}^m\} = \mathbb{0}
\quad\forall \, m$.
Na\"ively, this implies that there is an additional,
independent Yukawa coupling, $g_F$ for which
$\vec\beta_{A+B,F} =[\vec\gamma_{A+B}\gamma_F]\otimes[M_3]$.  
However, this is not the case as one parameter can be absorbed 
by a field-redefinition of the fermions. To see this note that
the two Yukawa terms proportional to  $g$ and $g_F$ can be expressed
as the real and imaginary part of one complex Yukawa, i.e., 
$g' e^{i \alpha \gamma_F}$ with $g = \cos\alpha \cdot g'$ 
and $g_F = i \sin\alpha \cdot g'$.
Applying the field redefinition  $\psi \to e^{-i \alpha \gamma_F/2} \psi$ 
absorbs the exponential term and yields the original Yukawa
in Eq.~\eqref{eq:Lag3d}, cf., Ref.~\cite{Roy:2012bor}.
}
To illustrate how  the larger Clifford algebra of
Eq.~\eq{Clif} is realized, we use the tensor decomposition of $\vec\gamma$ 
in terms of the spinor representation of $\SO(N_A\!+\!N_B)$
and the remaining two components
\begin{equation}
  \label{eq:gammadecomposition}
  \vec\gamma =\Big\{
  \overbrace{\underbrace{[\vec\gamma_{A+B}]\otimes[M_3]}_{N_A+N_B \;\mathrm{matrices}}}^{=\vec\beta_{A+B}},\,
  \overbrace{\underbrace{[\mathbb{1}_{A+B}]\otimes[M_{1,2}]}_{2\;\mathrm{matrices}}}^{\psi~\text{kinetic term}}
  \Big\}  \,.
\end{equation}
$\vec\gamma_{A+B}$ are square matrices of dimension $d_{A+B}=2^{\lfloor\frac{N_A\!+\!N_B}{2}\rfloor}$
that satisfy the $\SO(N_A\!+\!N_B)$ Clifford algebra. 
The square matrices $M_{i}$, with $i=1,2,3$, have the dimension 
$d_M = d_\gamma / d_{A+B}$ and must satisfy
\begin{equation}
  \{M_i, M_j\} = 2 \delta_{ij} \mathbb{1}_{d_M\times d_M}
\end{equation}
for the $\vec{\gamma}$ to fulfill Eq.~\eq{Clif}.
Note how it is essential to have a third ---independent from $M_{1,2}$---
element, $M_3$, that satisfies the $M_i$ Clifford algebra. 
Without it Eq.~\eq{Clif} could not be satisfied 
since $\{\vec{\gamma}_{A+B},\mathbb{1}_{A+B}\}\neq \mathbb{0}$.
The invariance under the internal $\SO(3)$ guarantees the 
existence of such an $M_3$.

Next we investigate $\SO(N_A)\times\SO(N_B)$--invariant GNY theories, i.e.,
theories with fermions coupled to two independent order parameters, that are
compatible with the enlarged $\SO(N_A\!+\!N_B)$ symmetry discussed above.

\subsection{\texorpdfstring{$\SO(N_A)\times\SO(N_B)$}{SO(NA) x SO(NB)} GNY Theory}

The $2+1$-dimensional GNY theory that is the main focus of the current work 
exhibits a global $\SO(N_A) \times \SO(N_B)$ symmetry under which 
two separate real order parameters $\vec{\Phi}_A$ and $\vec{\Phi}_B$, transform as vectors,
i.e., in the respective fundamental representation, with couplings to $d_\gamma$-component 
complex fermion fields. Compatibility with the enlarged $\SO(N_A\!+\!N_B)$ symmetry 
of the GNY theory in Eq.~\eq{Lag3d} places strong constraints on this less symmetric 
$\SO(N_A) \times \SO(N_B)$ model. The presence of the 
internal $\SO(3)$ ---typical for such Dirac systems in two space dimensions --- 
is unchanged from the $\SO(N_A\!+\!N_B)$ theory.

Explicitly, the $\SO(N_A) \times \SO(N_B)$--invariant GNY field theory of 
gapless~\footnote{Direct fermionic mass terms are compatible with the global symmetry but 
can be tuned to zero in order to study critical points.}
$d_\gamma$-component complex fermion fields 
$\psi$ is given by the Euclidean-signature Lagrangian
\begin{equation}\label{eq:Lag3d}
\begin{aligned}
  \Lagr_{2+1}^{A\times B }
  &=\frac{1}{2} (\partial_\mu \vec{\Phi}_A)^2
   -\frac{r_{A}}{2}(\vec{\Phi}_A)^2 
   -\frac{\lambda_A}{8} (\vec{\Phi}_A)^4\\
  &+\frac{1}{2} (\partial_\mu \vec{\Phi}_B)^2 
   - \frac{r_B}{2} (\vec{\Phi}_B)^2
   -\frac{\lambda_B}{8} (\vec{\Phi}_B)^4 \\
  &- \frac{\lambda_{AB}}{4} (\vec{\Phi}_A)^2 (\vec{\Phi}_B)^2 \\
  &+ \psi^{\dagger} \partial_\tau \psi 
  + \psi^{\dagger}\Big[\mathbb{1}_{\frac{d_\gamma}{2}\!\times\!\frac{d_\gamma}{2}} \otimes \sum_{k=1,2} \sigma_k \left(-i \partial _k\right)\Big] \psi \\
  &- \psi^{\dagger} \Big[g_A \vec{\Phi}_A\!\cdot\!\vec{\beta}_A 
  + g_B \vec{\Phi}_B\!\cdot\!\vec{\beta}_B\Big] \psi \,,
\end{aligned}
\end{equation}
with the two order parameters 
$\vec{\Phi}_A = \left(\Phi_{A,1}, \dots, \Phi_{A,{N_A}}\right)^T$ and 
$\vec{\Phi}_B = \left(\Phi_{B,1}, \dots, \Phi_{B,{N_B}}\right)^T$ represented 
by real scalar fields.
The parameters $r_{A,B}$, $\lambda_{A,B,AB}$ and $g_{A,B}$ are all real and 
are super-renormalizable in $2+1$ dimensions, i.e., their mass dimension is
$[r_{A,B}]=2$, 
$[\lambda_{A,B,AB}]=1$, 
$[g_{A,B}]=1/2$. However, in the $3+1$ continuation of the theory that we 
develop they are marginal and thus amenable to standard perturbation theory
methods via the  $\epsilon$-expansion approach. 

Note that we have suppressed allowed interactions
of the type $\vec\Phi^6$ and $\psi^\dagger\psi \vec\Phi^2$ that are marginal in $2+1$
since they are non-renormalizable in the continuation and thus their effect 
cannot be captured via the $\epsilon$-expansion. 
The fact that the $\epsilon$-expansion
cannot provide quantitative estimate of such effects is an intrinsic shortcoming 
of the approach.
UV/IR fixed
points of $d=3$ quartic Yukawa models featuring an $\psi^\dagger\psi
\vec\Phi^2$ interaction and scalar models with $\Phi^6$ contributions are still a field of active
research~\cite{Fraser-Taliente:2024rql,Kvedaraite:2025lgi}.

The scalar part of the Lagrangian in Eq.~\eq{Lag3d} is already compatible
with an emergent $\SO(N_A\!+\!N_B)$ symmetry with the order parameter
$\vec\Phi_{A+B} =(\vec\Phi_A, \vec\Phi_B)^T$ and
$\lambda_A=\lambda_B=\lambda_{AB}$. This is more subtle for the Yukawa-type
terms in Eq.~\eq{Lag3d} that provide the couplings of the order
parameters to the fermionic degrees of freedom in $\psi$. The $d_\gamma$
components of $\psi$ collectively summarize all degrees of freedoms related to
$\SO(N_A)$, $\SO(N_B)$, internal $\SO(3)$, and possible additional flavors.
Compatibility with the emergent $\SO(N_A\!+\!N_B)$ restricts the viable couplings of $\psi$
to the order parameters, i.e., the $d_\gamma\times d_\gamma$ matrices,
$\beta^a_A$ and $\beta^b_B$ with $a=1,\dots,N_A$ and $b=1,\dots,N_B$,
respectively. Put differently, compatibility constraints both the number of
fermionic components, $d_\gamma$, as well as the transformation properties of
the fermions under $\SO(N_A)\times \SO(N_B)$.

Concretely, there are three conditions on the fermionic components: 
{\itshape i)} 
The fermions must couple simultaneously to both the $\SO(N_A)$ and $\SO(N_B)$
order parameters, i.e., they must transform non-trivially under both, in order
for them to have the potential to stabilize an emergent $\SO(N_A\!+\!N_B)$
fixed point. 
{\itshape ii)} 
The number of fermionic components must be such that their couplings to both
the $\SO(N_{A/B})$ and the emergent $\SO(N_{A}\!+\!N_B)$ order parameters are
possible.
{\itshape iii)} 
The transformation properties under $\SO(N_{A})\times\SO(N_B)$ must permit the
fermions to satisfy the emergent Clifford algebra in Eq.~\eq{Clif}. 

These conditions can be automatically satisfied for any $\SO(N_{A/B})$ if the
fermions transform simultaneously in the spinor representation of both
$\SO(N_A)$ and $\SO(N_B)$.  Condition {\itshape i)} is then satisfied and the
$\SO(N_A)\times \SO(N_B)$--invariant interactions $\vec{\beta}_{A/B}$ admit a
tensor decomposition
\begin{equation}
  \vec\beta_A = \vec\gamma_A \otimes \widetilde{X}_B \otimes \widetilde{M}_A\,,\quad
  \vec\beta_B = \widetilde{X}_A \otimes \vec\gamma_B \otimes \widetilde{M}_B\,,
  \label{eq:betaABdecomposition}
\end{equation}
where $\vec\gamma_{A/B}$ and $\widetilde{X}_{A/B}$ are $d_{A/B}$-dimensional
matrices with $d_{A/B} = 2^{\lfloor N_{A/B}/ 2\rfloor}$. The $\vec{\gamma}_{A/B}$ satisfy the
individual Clifford algebras, $\{\gamma^m_{A/B},\gamma^n_{A/B}\}=2 \delta^{mn}
\mathbb{1}_{A/B}$.
The $d_{\widetilde{M}}$-dimensional matrices $\widetilde{M}_{A/B}$ with
$d_{\widetilde{M}} = d_\gamma / (d_{\gamma_A}  d_{\gamma_B})$ contain the
transformation under the internal symmetry as well as couplings among
additional copies of the multiplets, i.e., flavors. 
The invariance under spatial rotations in $d_s=2$
implies that $d_{\widetilde{M}}\geq 2$ (and a multiple of $2$).
This is always also compatible with imposing full Lorentz invariance.
In this case the internal spinor indices must be contracted with $\sigma_3$, e.g., 
for the minimal case $d_{\widetilde{M}}= 2$ this means that $\widetilde{M}_{A/B}=\sigma_3$.
There are two ways to implement the $\SO(N_{A/B})$ invariance of the Yukawas.
One either takes $\widetilde{X}_{A/B} = \mathbb{1}_{A/B}$ or, 
if $N_{A/B}$ is even, one can also use the additional Gamma matrix, $\gamma_{F,A/B}$,
that satisfies the respective Clifford algebra
and choose $\widetilde{X}_{A/B} = \gamma_{F,A/B}$.

To satisfy the condition {\itshape iii)}, we construct 
the $\vec{\beta}_{A+B}$ elements of the enlarged Clifford algebra by 
taking the $N_A+N_B$ elements of $\vec{\beta}_A$ and $\vec{\beta}_B$ to be
\textit{linearly independent}. To this end they must satisfy the
{\itshape compatibility condition}~\cite{PhysRevResearch.2.022005}
\begin{equation}
  \label{eq:betacom}
  \{\beta^a_A,\beta^b_B\} =\mathbb{0}_{d_\gamma \times d_\gamma}\,.
\end{equation}

Satisfying condition {\itshape ii)}, on the other hand, requires
\begin{equation}\label{eq:dMtilde}
  d_M \times d_{A+B} = d_\gamma = d_{\widetilde{M}}\times d_A \times d_B\,.
\end{equation}
Because of the floor function in the dimension of the spinor representation and
the various options to choose $\widetilde{X}_{A/B} \in \{\mathbb{1}_{A/B},
\gamma_{F,A/B}\}$, we distinguish two cases.
\begin{itemize}
  \item[{\itshape a)}] {\itshape $N_A$ and $N_B$ both odd.}\\
  In this case $d_{A+B} = 2 \times d_A\times d_B$, thus
  $d_M=d_{\widetilde{M}}/2$, and the only Yukawas that are invariant have
  $\widetilde{X}_{A/B} = \mathbb{1}_{A/B}$.
  Satisfying the compatibility condition in Eq.~\eqref{eq:betacom} is thus 
  only possible if 
  \begin{equation}
  \label{eq:Mtildeanticom}
  \{\widetilde{M}_A,\widetilde{M}_B\}=\mathbb{0}\,.
  \end{equation}
  In the minimal case $d_M=2$, we have $d_{\widetilde{M}}=4$. As a consequence,
  there are enough fermionic components to ensure both the compatibility with
  $\SO(N_A\!+\!N_B)$ and simultaneously preserve Lorentz invariance in the
  Yukawas by choosing $\widetilde{M}_A = \sigma_1 \otimes \sigma_3$ and
  $\widetilde{M}_B = \sigma_2 \otimes \sigma_3$ with the $\sigma_3$s' acting on
  the internal Dirac indices.
  \item[{\itshape b)}] {\itshape $N_A$ and $N_B$ both even or one even and one odd.}\\
  In this case $d_{A+B} = d_A\times d_B$, thus $d_M=d_{\widetilde{M}}$, and
  contrary to the previous case, there are two possibilities for invariant
  Yukawas $\widetilde{X}_{A/B} \in \{\mathbb{1}_{A/B},\gamma_{F,A/B}\}$ for
  $N_{A/B}$ even with:
  \begin{equation}
    \{\gamma_{F,A/B},\vec\gamma_{A/B}\}=\mathbb{0}\,.
  \end{equation}
  This freedom opens the possibility to have different ways to 
  implement the compatibility condition. 
  Concretely, if both $\widetilde{X}_{A/B} = \mathbb{1}_{A/B}$ or both are 
  $\widetilde{X}_{A/B} = \gamma_{F,A/B}$ then the 
  compatibility condition can only be satisfied via the $\widetilde{M}_{A/B}$ 
  by means of Eq.~\eqref{eq:Mtildeanticom}. 
  In this setup, 
  the minimal case $d_M=d_{\widetilde{M}}=2$, is incompatible 
  with Lorentz invariance 
  as this necessitates that $\widetilde{M}_A=\widetilde{M}_B=\sigma_3$
  in contrast to Eq.~\eqref{eq:Mtildeanticom}.
  Instead, if $\widetilde{X}_{A} = \mathbb{1}_{A}$ and $\widetilde{X}_{B}=  \gamma_{F,B}$
  or vice versa, then the compatibility condition is automatically satisfied 
  as long as 
  \begin{equation}
  \label{eq:Mtildecom}
  [\widetilde{M}_A,\widetilde{M}_B]=\mathbb{0}\,.
  \end{equation}
  Thus, in this setup the minimal case $d_M=d_{\widetilde{M}}=2$ is compatible
  with Lorentz invariance by choosing $\widetilde{M}_{A/B}=\sigma_3$.
\end{itemize}

For illustration, consider the physically relevant case on the honeycomb lattice of graphene, describing $d_\gamma = 8$ fermionic degrees
of freedom with Yukawa interactions~\cite{Herbut:2023xgz}.
In general, the honeycomb lattice admits 56 distinct pairwise anticommuting order parameters~\cite{PhysRevB.80.205319}. Out of the 56, for example, the $(N_A =3)$-component antiferromagnetic Néel order parameter and the $(N_B = 2)$-component Kekulé valence-bond solid order parameter take the form
\begin{equation}\label{eq:AFMVBS}
  \beta_A^i = \sigma_i \otimes \sigma_3 \otimes \sigma_3\,,\quad
  \beta_B^j = \mathbb{1}_{2\times 2} \otimes \sigma_j \otimes \sigma_3\,,
\end{equation}
with $i=1,2,3$ and $j=1,2$. 
Thus, this theory corresponds to an example of the second setup in case {\itshape b)} above.
Taking $\widetilde{X}_B=\sigma_3$ allows to implement Lorentz invariance, i.e., $\widetilde{M}_A=\widetilde{M}_B=\sigma_3$,
without increasing the number of fermionic degrees of freedom.
Lorentz invariance is also possible in the first setup of case {\itshape b)},
i.e., $\widetilde{X}_{A/B}=\mathbb{1}_{A/B}$, albeit only by 
doubling the fermionic degrees of freedom to $d_\gamma = 16$ such that
\begin{equation}
  \beta_{A}^i = \sigma^i \otimes \mathbb{1}_B \otimes \widetilde{M}_A\,,\quad
  \beta_{B}^j = \mathbb{1}_A \otimes \sigma^j \otimes \widetilde{M}_B\,,
\end{equation}
with, e.g., $\widetilde{M}_A = \sigma_1 \otimes \sigma_3$ and $\widetilde{M}_B
= \sigma_2 \otimes \sigma_3$.

Having identified the first $N_A+N_B$ elements of the Clifford algebra in Eq.~\eqref{eq:Clif},
we turn to the remaining two elements associated to rotations in the two spatial dimensions.
For Yukawa interactions that are manifestly Lorentz invariant, i.e., 
spinor indices contracted with $\sigma_3$, the remaining two elements
are directly obtained from the (Euclidean-signature) 
kinetic term in Eq.~\eqref{eq:Lag3d}, namely:
\begin{equation}
  [\mathbb{1}_{\frac{d_\gamma}{2} \times \frac{d_\gamma}{2}}]
  \otimes
  [-i \sigma_3 \sigma_{1,2}]\,.
\end{equation}
They complete the enlarged Clifford algebra and lead to a kinetic term 
for the fermions that is both $\SO(N_A)\times\SO(N_B)$- and Lorentz-invariant.\\

We have thus shown that in $d=2+1$ all $\SO(N_A)\times\SO(N_B)$-invariant
theories in Eq.~\eqref{eq:Lag3d} that are compatible with an emergent $\SO(N_A+N_B)$ 
{\itshape and} contain sufficient fermionic components to satisfy the 
enlarged Clifford algebra in Eq.~\eq{Clif} have Yukawa interactions that can be 
chosen to be manifestly Lorentz-invariant.
For this reason, the Lorentz-invariant $4-\epsilon$ theory that we discuss in 
the next section is expected to provide a valid continuation of this class of 
Lorentz-invariant $d=3$ theories.

\section{\texorpdfstring{GNY continuations to $\boldsymbol{d=4-\epsilon}$}{GNY continuations to d=4-epsilon}\label{sec:model4d}}

\begin{table}[t]
\begin{tabular*}{\linewidth}{@{\extracolsep{\fill} } l c c c}
  \hline\hline\\[-0.75em]
  & $\SO(N_A)$     & $\SO(N_B)$     & $\mathrm{U}(N_\Psi)$\\ \hline\\[-0.75em]
$\Psi$         & Spinor         & Spinor         & $\mathbf{N_\Psi}$\\
$\vec{\Phi}_A$ & $\mathbf{N_A}$ & --             & --\\
$\vec{\Phi}_B$ & --             & $\mathbf{N_B}$ & --\\
\hline
\end{tabular*}
\caption{
 Field content of the Lorentz-invariant $(4-\epsilon)$-dimensional theory in
 Eq.~\eq{bareLagr4d} along with transformation properties of the fields 
 under the global $\SO(N_A) \times \SO(N_B)\times \U(N_\Psi)$ symmetry. 
 The global $\U(N_\Psi)$ symmetry is a flavor symmetry under which the $N_\Psi$ number of
 four-component Dirac fermions, $\Psi$, transform. It is employed to
 facilitate the continuation to $d=2+1$. \label{table:quantities4d}}
\end{table}

Next, we develop the manifestly Lorentz-invariant dimensional continuation of 
Eq.~\eq{Lag3d} to $d=4-\epsilon$ spacetime dimensions.
The theory exhibits an $\SO(N_A) \times \SO(N_B) \times \mathrm{U}(N_\Psi)$ 
global symmetry, featuring scalar fields $\vec{\Phi}_{A,B}$ in the 
fundamental representation of the respective orthogonal subgroups $\SO(N_{A,B})$, 
as well as a four-component Dirac fermion, $\Psi$, that transforms in
the spinor representation of both $\SO(N_A)$ and $\SO(N_B)$.
Additionally, $\Psi$ transforms in the fundamental representation of an
$\U(N_\Psi)$ flavor symmetry, i.e., $\Psi$ contains $N_\Psi$ four-component
Dirac fermions. Imposing the $\U(N_\Psi)$ flavor symmetry restricts
the allowed interactions between the $N_\Psi$ fermions and the scalar fields.
This facilitates a direct parameter mapping between the $d=3+1$ and the
$d=2+1$ theory from Sec.~\ref{sec:modeld3}.
\autoref{table:quantities4d} summarizes the field content and its symmetry properties.

In Minkowskian signature, the corresponding renormalizable Lagrangian reads
\begin{equation}\label{eq:bareLagr4d}
\begin{aligned}
\Lag_{3+1}^{A\times B}  &= \frac{1}{2}(\partial_\mu \vec{\Phi}_A)^2 
             - \frac{r_A}{2}(\vec{\Phi}_A)^2 
             - \frac{\lambda_A}{8} (\vec{\Phi}_A\!\cdot\!\vec{\Phi}_A)^2\\  
            &+ \frac{1}{2}(\partial_\mu \vec{\Phi}_B)^2
             - \frac{r_B}{2}(\vec{\Phi}_B)^2 
             - \frac{\lambda_B}{8} (\vec{\Phi}_B\!\cdot\!\vec{\Phi}_B)^2 \\
            &- \frac{\lambda_{AB}}{4} (\vec{\Phi}_A)^2 (\vec{\Phi}_B)^2 + \overline{\Psi} i \partial_\mu \gamma^\mu \Psi + \Lagr_{\mathrm{yuk}}\,,
\end{aligned}
\end{equation}
with $\mu = 0,\dots,3$, $\gamma^\mu$ the $4\times 4$ Dirac matrices 
satisfying $\{\gamma^\mu,\gamma^\nu\}=2 g^{\mu\nu} \mathbb{1}_{4\times 4}$ and
$\overline{\Psi} \equiv \Psi^\dagger \gamma_0$.
The mass dimensions of the parameters in the $3+1$ theory
are $[r_{A,B}]=2$ and $[\lambda_{A,B,AB}]=0$. 

The most general, renormalizable coupling of the
fermions to the order parameters is a Yukawa-type interaction, $\Lagr_{\mathrm{yuk}}$, 
that depends on two complex couplings $y_{A,B}$ with mass dimension $[y_{A,B}]=0$.
This is best seen if first written in terms of the two-component chiral Weyl 
spinors contained in the Dirac fermion $\Psi =(\xi,\,\eta^\dagger)^T$.
To this end we explicitly write the {\itshape global-symmetry indices}
$\xi_{ijk}$ and $\eta^{ijk}$ where 
$i=1,\dots,2^{\left\lfloor N_A/2\right\rfloor}$,  
$j=1,\dots,2^{\left\lfloor N_B/2 \right\rfloor}$, and 
$k =1,\dots,N_\Psi$. 
A lowered index indicates that the field transforms in the 
spinor and fundamental representation of $\SO(N_{A/B})$ and $\U(N_\Psi)$, respectively, while
a raised one indicates that it transforms in the respective conjugate representation.
The most general Yukawa interaction respecting Lorentz and the global symmetry then reads
\begin{align}\label{eq:YukNaive}
\Lagr_{\mathrm{yuk}} & = -\; y_A \,\eta^{ijk} ( \vec{\Phi}_A \cdot \vec{\gamma}_{A} )_{i \phantom{i'}}^{\phantom{i} i'} (\widetilde{X}_B)_{j \phantom{j'}}^{\phantom{j} j'} \xi_{i' j' k} \nonumber \\
    &\phantom{ = \; }- y_B \,\eta^{ijk} ( \vec{\Phi}_B \cdot \vec{\gamma}_{B})_{j \phantom{j'}}^{\phantom{j} j'} (\widetilde{X}_A)_{i \phantom{i'}}^{\phantom{i} i'} \xi_{i' j' k} + \mathrm{h.c.}  \,,
\end{align}
with $y_{A/B}$ complex and $\vec{\gamma}_{A/B}$ satisfying the individual
Clifford algebras for $\SO(N_{A/B})$, i.e.,
$\{\gamma^m_{A/B},\,\gamma^n_{A/B}\} = 2\, \delta^{mn} \, \mathbb{1}_{A/B}$.
$\SO(N_{A/B})$ invariance implies that $\widetilde{X}_{A/B}=\mathbb{1}_{A/B}$ for $N_{A/B}$ odd, 
or $\widetilde{X}_{A/B} \in \{\mathbb{1}_{A/B}, \gamma_{F,A/B}\}$ for $N_{A/B}$ even, cf., 
the discussion in Sec.~\ref{sec:modeld3} below  Eq.~\eq{betaABdecomposition}.
Rewriting Eq.~\eq{YukNaive} in terms of four-component Dirac fermions and  suppressing all indices yields
\begin{equation}\label{eq:Yuk4DComplex}
\begin{aligned}
\Lagr_{\mathrm{yuk}} = 
   & -\overline{\Psi}\left\{ \left[\mathrm{Re}(y_A) + i\mathrm{Im}(y_A) \gamma_5 \right] \left(\vec{\Phi}_A \cdot \vec{\gamma}_{A}\right) \widetilde{X}_B \right\} \Psi \\
   & -\overline{\Psi}\left\{ \left[\mathrm{Re}(y_B) + i\mathrm{Im}(y_B) \gamma_5 \right] \left(\vec{\Phi}_B \cdot \vec{\gamma}_{B}\right) \widetilde{X}_A \right\} \Psi \,.
\end{aligned}
\end{equation}
with $\gamma_5$ acting on the four-component Dirac fermions.
We see that ---apart from the different options for $\widetilde{X}_{A/B}$ that we discuss below---
the Yukawas in $d=3+1$ theory contain two complex parameters $y_A$ and $y_B$, whereas 
the $d=2+1$ theory only two real parameters $g_A$ and $g_B$.

To make the connection to the $2+1$ theory in Eq.~\eq{Lag3d}, we express the Yukawas in Eq.~\eq{Yuk4DComplex}
via their tensor decomposition
\begin{equation*}\label{eq:Yuk4dMassOps}
  \Lagr_{\mathrm{yuk}} = -\!\!\sum_{X=A,B}\!\! \Psi^\dagger\!\!  \left[\mathrm{Re}(y_X) \vec{\beta}_{X,\mathrm{Re}} + \mathrm{Im}(y_X) \vec{\beta}_{X,\mathrm{Im}} \right] \cdot\vec{\Phi}_X  \Psi\,,
\end{equation*}
with
\begin{equation}
  \begin{split}
  \vec{\beta}_{A, \mathrm{Re}} & = \vec{\gamma}_A \otimes \widetilde{X}_B   \otimes \mathbb{1}_{N_\Psi} \otimes \gamma_0 \,,          \\ 
  \vec{\beta}_{A, \mathrm{Im}} & = \vec{\gamma}_A \otimes \widetilde{X}_B   \otimes \mathbb{1}_{N_\Psi} \otimes i\gamma_0 \gamma_5 \,,  \\[0.5em]
  \vec{\beta}_{B, \mathrm{Re}} & = \widetilde{X}_A   \otimes \vec{\gamma}_B \otimes \mathbb{1}_{N_\Psi} \otimes \gamma_0 \,,          \\
  \vec{\beta}_{B, \mathrm{Im}} & = \widetilde{X}_A   \otimes \vec{\gamma}_B \otimes \mathbb{1}_{N_\Psi} \otimes i\gamma_0 \gamma_5 \,, 
  \end{split}
    \label{eq:Betas4d}
\end{equation}
cf., Eq.~\eq{betaABdecomposition}.
In this schematic $M_A \otimes M_B \otimes M_{N_\Psi} \otimes M_{L}$ form,
$M_A$, $M_B$ and $M_{N_\Psi}$ act on the respective representation indices of
$\SO(N_A)$, $\SO(N_B)$ and $\U(N_\Psi)$, while $M_{L}$ acts on the Lorentz
spinor indices of the Dirac fermion.
$\widehat{M}_{A/B} \equiv M_{N_\Psi}\otimes M_L$ are square matrices of dimension
$d_{\widehat{M}}=4N_\Psi$ and correspond to the matrices $\widetilde{M}_{A/B}$ 
in the $d=2+1$ theory (see Eq.~\eq{betaABdecomposition}).

By appropriately choosing the generic Yukawas in Eq.~\eqref{eq:Betas4d} 
we shall define the continuation of all the $2+1$ models presented in the
previous section.
We proceed analogously and discuss in turn two cases:
\begin{itemize}
\item[{\itshape a)}] {\itshape $N_A$ and $N_B$ both odd.}\\
Here, both $\widetilde{X}_A$ and $\widetilde{X}_B$ can only take the value
$\mathbb{1}_{A/B}$. Thus, in total analogy to Eq.~\eqref{eq:Mtildeanticom},
compatibility implies that
\begin{equation}
\label{eq:Mhatanticom}
\{\widehat{M}_A,\widehat{M}_B\}=\mathbb{0}\,.
\end{equation}
This is only possible if one Yukawa coupling is real while the 
other purely imaginary, e.g., 
\begin{align}
  \phantom{M}\{ \vec\beta_{A, \mathrm{Re}}, \vec\beta_{B, \mathrm{Im}}\} &
  = 8\,\vec\gamma_A\otimes \vec\gamma_B \otimes\mathbb{1}_{N_\Psi}\! 
  \otimes \{ \gamma_0, i\gamma_0 \gamma_5 \} \nonumber\\
  &= \mathbb{0}\quad\text{since}\quad \{\gamma_0,\gamma_5\}=\mathbb{0}\,.
\end{align}
Instead, if both Yukawas were either real or purely imaginary the
anticommutator would not vanish
\begin{equation*}
  \{ \vec\beta_{A, \mathrm{Re}}, \vec\beta_{B, \mathrm{Re}}\} \neq \mathbb{0}\,,\quad
  \{ \vec\beta_{A, \mathrm{Im}}, \vec\beta_{B, \mathrm{Im}}\} \neq \mathbb{0}\,,
\end{equation*}
and the continuation would describe models  with {\itshape incompatible} order parameters, 
see, e.g., Refs.~\cite{Herbut:2022zzw,Uetrecht:2023uou,Herbut:2023xgz} for recent studies.

Therefore, in this case the $3+1$ Lagrangian in Eq.~\eq{bareLagr4d} is 
compatible with the emergence of an $\SO(N_A\!+\!N_B)$ symmetry if
\begin{equation}\label{eq:compatible-4d}
\begin{aligned}
  \mathrm{Re}(y_A) &= g_A\,,  &\mathrm{Im}(y_A) &= 0\,,\\
  \mathrm{Re}(y_B) &= 0\,,    &\mathrm{Im}(y_B) &= g_B\,,
\end{aligned}
\end{equation}
or the analogous case with $A\leftrightarrow B$.

\item[{\itshape b)}] {\itshape $N_A$ and $N_B$ both even or one even and one odd.}\\
For $N_{A/B}$ even the options of choosing 
$\widetilde{X}_{A/B} \in \{\mathbb{1}_{A/B},\gamma_{F,A/B}\}$ opens the 
possibility to construct different models that are compatible.
If both $\widetilde{X}_{A/B} = \mathbb{1}_{A/B}$ or both are 
$\widetilde{X}_{A/B} = \gamma_{F,A/B}$ then the 
compatibility condition can only be satisfied via the 
$\widehat{M}_{A/B}$ by means of Eq.~\eqref{eq:Mhatanticom}.
Thus the continuation of these models is as in case {\itshape a)}, 
i.e., one real and one imaginary coupling (Eq.~\eqref{eq:compatible-4d}).
Instead, if $\widetilde{X}_{A} = \mathbb{1}_{A}$ and $\widetilde{X}_{B}=  \gamma_{F,B}$
or vice versa, then the compatibility condition is automatically satisfied 
as long as 
\begin{equation}
\label{eq:Mhatcom}
[\widehat{M}_A,\widehat{M}_B]=\mathbb{0}\,.
\end{equation}
Since $\gamma_0^2 = \gamma_5^2= \mathbb{1}$ this implies
that the Yukawas in the continuation of these models are
either both real or purely imaginary, i.e., either
\begin{equation}\label{eq:compatible2-4d}
\begin{aligned}
  \mathrm{Re}(y_A) &= g_A\,, &\mathrm{Im}(y_A) &= 0\,,\\
  \mathrm{Re}(y_B) &= g_B\,, &\mathrm{Im}(y_B) &= 0\,,
\end{aligned}
\end{equation}
or the analogous expressions with $\mathrm{Re}\leftrightarrow \mathrm{Im}$.
We note that the distinct parameter subspaces of the full model
with complex Yukawas are protected by a CP symmetry.
For example, if the coupling are those in  Eq.~\eqref{eq:compatible-4d},
$\vec{\Phi}_A$ transforms as a scalar and $\vec{\Phi}_B$ as a pseudoscalar, 
whereas if Eq.~\eqref{eq:compatible2-4d} holds, both are scalars.
This means that the renormalization group cannot induce Yukawas from different subspaces.
\end{itemize}

The final step in defining the dimensional continuation is the mapping of 
the number of fermion degrees of freedom between the $d=4-\epsilon$ and the $d=3$
theory. 
The multiplicity associated to the global $\SO(A/B)$ symmetries is trivial
since the fermions transform under the same representations in the
$d=4-\epsilon$ and $d=3$ theory.
Thus, only the counting associated to internal and flavor degrees of freedom
can be non-trivial.
When employing the $\epsilon$-expansion in fermionic 
theories (see, e.g., Ref.~\cite{Uetrecht:2023uou}) it is standard to use 
the number of flavors in the $4-\epsilon$ theory, $N_\Psi$, to interpolate the
degrees of freedom across dimensions, i.e.,
\begin{equation}
  \label{eq:Nmap}
  4 \times N_\Psi = d_{\widehat{M}}  \quad\overset{4d\to 3d}{=}\quad d_{\widetilde {M}}\,.
\end{equation}
Equivalently, using the definition of $N_f$ below Eq.~\eqref{eq:compatible} 
and Eq.~\eqref{eq:dMtilde} this can be cast into the relation
\begin{equation}
\label{eq:NPsitoNf}
   N_\Psi \overset{4d\to 3d}{=}N_f / d_A / d_B\,.
\end{equation}

The $4-\epsilon$ theory is per construction Lorentz invariant.
Therefore, it provides a valid continuation only to the $2+1$ theories 
for which Lorentz invariance is manifest.
As extensively discussed in Sec.~\ref{sec:modeld3}, it is always
possible to ensure that the $d=3$ theory is both compatible with the emergent $\SO(N_A\!+\!N_B)$ 
and Lorentz invariant by including sufficient fermionic components.
The {\itshape minimal} number of fermionic components for which this is possible depends 
on the details of the Yukawas, i.e., on $\widetilde{X}_{A/B}$.
Following the findings from Sec.~\ref{sec:modeld3} on the minimal degrees of
freedom required for compatibility with $\SO(N_A\!+\!N_B)$ and Lorentz invariant
leads to the mapping
\begin{equation}
\label{eq:NPsimin}
  N_\Psi^{\mathrm{min}} \overset{4d\to3d}{=}
    \begin{cases}
      1\, & \text{for models with}~ \{\widetilde{M}_A, \widetilde{M}_B\} =\mathbb{0}\,, \\
      \frac{1}{2}\, & \text{for models with}~ ~[\widetilde{M}_A, \widetilde{M}_B ] =\mathbb{0}\,.
    \end{cases}
\end{equation}

This concludes the construction of the $4-\epsilon$ theories that provide a valid dimensional continuation 
to all the $d=3$ theories discussed that are compatible with an emergent $\SO(N_A\!+\!N_B)$ symmetry, 
satisfy the Clifford algebra in Eq.~\eqref{eq:Clif}, and are manifestly Lorentz invariant.
In the next section, we compute the RGEs and assess the stability of the appearing fixed points
in the $d=3$ theories within the framework of the $\epsilon$ expansion.

\section{Analysis of critical behavior\label{sec:analysis}}

We have computed the renormalization group equations (RGEs) of the Lagrangian
in Eq.~\eq{bareLagr4d} in $d=4-\epsilon$ and the $\overline{\rm MS}$ scheme at
three-loop accuracy using the \texttt{FoRGEr} toolkit~\cite{Steudtner:FoRGEr},
which uses the three-loop expressions from
Refs.~\cite{Steudtner:2021fzs,Jack:2023zjt,Steudtner:2024teg}.
We obtain the $\beta$ functions of the two Yukawa couplings,
\begin{align}
  \beta_{y_X}(\epsilon)\equiv\frac{d y_X}{d\log \mu} \,,
\end{align}
with $X\in\{A,\,B\}$, as well as the $\beta$ functions of the three quartic couplings,
\begin{align}
  \beta_{\lambda_X}(\epsilon)\equiv\frac{d \lambda_X}{d\log \mu} \,,
\end{align}
with $X\in\{A,\,B,\,AB\}$.
We have also computed the anomalous dimensions of the masses of the two order
parameters $r_{A,B}$, along with the anomalous field dimensions of the fermions
and order parameters,
\begin{align}
 \gamma_{r_X} \equiv
  \frac{d \log r_{X}}{d\log\mu} \,,\ 
  \gamma_\Psi \equiv
  \frac{d \log Z^{1/2}_\Psi}{d\log\mu} \,, \  \gamma_{\Phi_X} \equiv
  \frac{d \log Z^{1/2}_{\Phi_{X}}}{d\log\mu}  \label{eq:ADM_order_param} \,, 
\end{align}
where $X\in\{A,\,B\}$. The $Z$s denote renormalization constants related to the
bare fields in the convention $\Psi^{(0)} = Z_\Psi^{1/2} \Psi$ and
$\vec{\Phi}_X^{(0)} = Z_{\Phi_X}^{1/2} \vec{\Phi}_X$.

We have cross-checked our results with the one-loop results of
Ref.~\cite{PhysRevB.97.041117} for arbitrary values of $N_{A,B}$, with the
two-loop computation of Ref.~\cite{Uetrecht:2023uou} for $N_A = N_B = 3$, and
with Ref.~\cite{Zerf:2017zqi} in which the chiral Ising, XY, and Heisenberg
model, comprised of one single real order parameter with $N_{A/B} = 1,2,3$,
were computed at four-loop order. 
Additionally, we verified our results against a one-loop computation for the
special case $N_A = 1$ and $N_B = 2$, which describes the $\mathbb{Z}_2 \times
\grO(2)$ biconal fixed point of a GNY theory with three scalar
fields~\cite{PhysRevB.84.113404}.

The full $\beta$ functions and anomalous dimensions for general values of
$N_{A,B}$, $N_\Psi$, and complex $y_{A,B}$ are provided as a 
\texttt{Mathematica} file in the attachment and in the ancillary files of 
the arXiv submission of this paper.

To investigate the critical behavior of the model we determine the fixed points
of the $\beta$ functions in $d=4-\epsilon$, i.e., by determining the critical
points $c_m^{*} \in \{ g_A^{*2}, g_B^{*2}, \lambda_A^*, \lambda_B^*,
\lambda_{AB}^* \}$ for which the $\beta$ functions vanish.  The stability of a
given fixed point with coordinates $c_m^{*}$ is then found by linearizing the
RG flow in its vicinity, via
\begin{align}
	\beta_{c_{m}} = -S_{m n} \left( c_{n} - c_n^{*} \right) + \mathcal{O}\left( (c - c^{*})^2 \right)\,,
\end{align}
where
\begin{align}\label{eq:stabmat}
S_{m n} = -(\partial \beta_{c_m} / \partial c_n) \big\vert_{c^{*}}\,,
\end{align}
denotes the \textit{stability matrix} with five eigenvalues $\theta_k$.  The
sign of the eigenvalues determines the orientation of the RG flow: positive
values of $\theta_k$ indicate an IR repulsive (unstable, relevant) direction,
while negative values correspond to an IR attractive (stable, irrelevant)
direction. 
A fixed point is considered stable when all directions in the RG flow are IR
attractive. 
Throughout our analyses, we tune the bosonic masses $r_A$ and $r_B$ to be zero,
and we note that they correspond to relevant directions in the RG flow.

The remainder of this section is structured as follows.
Firstly, we investigate that models that are compatible with an 
emergent $\SO(N_A+N_B)$ symmetry and analyze their isotropic fixed point (IFP) 
at the three-loop order.  
By evaluating its stability in $d=3$, i.e., in the limit $\epsilon \to 1$, for
different values of $N_A$, $N_B$, and $N_\Psi$, we shall determine whether the
models can exhibit an emergent $\SO(N_A\!+\!N_B)$ symmetry with a stable IFP. 
Secondly, we utilize our general three-loop results to revisit the incompatible
$\SO(3) \times \SO(3)$ GNY model from Refs.~\cite{PhysRevLett.128.117202,Herbut:2022zzw,
Uetrecht:2023uou}. 

\subsection{Compatible Models for general \texorpdfstring{$N_A, N_B$}{NA, NB} and \texorpdfstring{$N_f$}{Nf}\label{subsec:compatible}}

In the following, we search for fixed-point solutions of all the {\itshape
compatible} $\SO(N_A)\times \SO(N_B)$ models in $d=3$ discussed in
Sec.~\ref{sec:modeld3} by employing their $d=4-\epsilon$ continuations of
Sec.~\ref{sec:model4d}. 
Per construction all models have the potential to satisfy the enlarged Clifford
algebra in Eq.~\eqref{eq:Clif} and satisfy the compatibility condition of
Eq.~\eqref{eq:betacom}.

As discussed in the previous sections, for generic $N_{A/B}$ there are multiple 
(inequivalent) ways to implement compatibility with manifest Lorentz invariance:
\begin{description}
  \item[Type I] Models with $\{\widetilde{M}_A,\widetilde{M}_B\}=\mathbb{0}$, i.e., 
    Eqs.~\eqref{eq:Mtildeanticom} and \eqref{eq:Mhatanticom}, in which the $4-\epsilon$ Yukawas 
    ($y_A$, $y_B$) are continued to the $d=3$ couplings ($g_A$, $g_B$) via Eq.~\eqref{eq:compatible-4d}.
  \item[Type II] Models with $[\widetilde{M}_A,\widetilde{M}_B]=\mathbb{0}$, i.e., 
    Eqs.~\eqref{eq:Mtildecom} and \eqref{eq:Mhatcom}, in which the $4-\epsilon$ Yukawas 
    ($y_A$, $y_B$) are continued to the $d=3$ couplings ($g_A$, $g_B$)
    via Eq.~\eqref{eq:compatible2-4d}.
\end{description}

By explicitly computing the $\beta$ functions and anomalous dimensions
of the two types of models at the three-loop level we have uncovered that
they are all related via
\begin{align*}
  \beta^{\text{Type\,I}}(N_\Psi)  &= \beta^{\text{Type\,II}}(N_\Psi)\,,\\
  \gamma^{\text{Type\,I}}(N_\Psi) &= \gamma^{\text{Type\,II}}(N_\Psi)\,,
\end{align*}
at three-loop order. Therefore, in what follows we shall not distinguish between 
type I and type II models. Their only difference is that the minimal number 
of fermionic components is different, see $N_\Psi^{\text{min}}$ in Eq.~\eqref{eq:NPsimin}.

We are only interested in the IFP, at which
the system exhibits an emergent $\SO(N_A\!+\!N_B)$ symmetry. 
Therefore, we determine the IFP in the isotropic subspace
\begin{equation}
  g \equiv g_A = g_B, \quad \lambda \equiv \lambda_A = \lambda_B = \lambda_{AB}\,.  
\end{equation}
Having determined the isotropic IFP we then assess its stability in $d=3$ using {\itshape the full $\beta$ functions} 
of the compatible model in Eq.~\eq{bareLagr4d} by taking the limit $\epsilon \to 1$.
The IFP depends solely on the variables of  the $\SO(N_A\!+\!N_B)$ theory, i.e., on 
\begin{equation}
  N \equiv N_A + N_B\,,\quad  N_f \equiv d_\gamma / 4\,,
\end{equation}
with the relation between $N_\Psi$ and $N_f$ given in Eq.~\eqref{eq:NPsitoNf} (see also Eq.~\eqref{eq:compatible}).

\begin{widetext}
In the isotropic subspace, the $\beta$ functions of the compatible model at three-loop order then take the form
	\begin{align}
		\beta_{g^2} =-\epsilon g^2
		  & - \frac{1}{16 \pi^2} 2 g^4 (N - 4 - 2N_f) \notag                                                                                                                       \\
		  & + \frac{1}{(16 \pi^2)^2} \frac{g^2}{2} \Big[ - (64 + 9 N^2  - 64 N  + 48 N_f) g^4 - 8  (2 + N) g^2 \lambda + (2 + N) \lambda^2 \Big]   \notag                                                                                                    \\
		  & + \frac{1}{(16 \pi^2)^3} \frac{g^2}{8} \Big[ 48  (2 + N) (8 - N + 5 N_f) g^4 \lambda +  (2 + N) (92 - N - 30 N_f) g^2 \lambda^2 - (2 + N) (8 + N) \lambda^3 \nonumber                                                        \\
		  & \qquad+ \Big( 
    32 \big( 40 + 10 N_f + N_f^2 + 36 (2 + N_f) \zeta_3 \big) 
    - 16 N \big( 164  + 17 N_f  - 16 N_f^2 + 16 \big( 9 + N_f \big) \zeta_3 \big) 
    \nonumber                                         \\
		  & \qquad
    + N^2 \big( 756  + 86 N_f + 96 (3 -  N_f) \zeta_3 \big)  + N^3 (-109 + 48 \zeta_3)   \Big) g^6  \Big] \label{eq:betag2Compatible}\,,          
\end{align}
\begin{align}
		\beta_{\lambda}=-\epsilon  \lambda
		  & + \frac{1}{16 \pi^2} \left[-16  N_f g^4 + 8  N_f g^2 \lambda + (8 + N) \lambda^2 \right]\notag                                                                          \\
		  & +\frac{1}{(16 \pi^2)^2} \Big[ 128  N_f g^6 + 4 (12 - 5 N) N_f g^4  \lambda - 4 (8 + N) N_f g^2  \lambda^2 - 3 (14 + 3 N) \lambda^3) \Big] \nonumber \\
		  & + \frac{1}{(16 \pi^2)^3} \frac{1}{8} \Big[  \big( 2960 + 922 N + 33 N^2  + 96 (22 + 5 N) \zeta_3 \big) \lambda^4 + 4 (322 + 65 N) N_f g^2  \lambda^3  \nonumber                                         \\
		  & \qquad+ 4  N_f \big( 208 + 836 N + 39 N^2  - 192 N_f - 24 N N_f  + (2208 - 24 N (10 + N)) \zeta_3 \big) g^4 \lambda^2  \nonumber                                                                                   \\  
        & \qquad+ 4  N_f  \big( -928 N - 11 N^2  + 312 N N_f   - 48 N (-14 + N)  \zeta_3 - 16 (216 - 89 N_f + 156 \zeta_3) \big) g^6 \lambda
    \nonumber                                    \\
		  & \qquad+ 16 N_f \big( -27 N^2 + 12 N (96  + N_f)  - 128 (2 + 5 N_f + 3 \zeta_3) \big) g^8  \Big] \label{eq:betalamCompatible} \,.                                                                 
	\end{align}
For completeness and later use, we collect the anomalous dimensions of the
scalar and fermion field as well as the one of the squared boson mass $r = r_{A,B}$ in
App.~\ref{app:anomDimComp}.
\end{widetext}

We note that Eqs.~\eq{betag2Compatible} and~\eq{betalamCompatible}
extend the results obtained in Refs.~\cite{PhysRevB.96.165133,Zerf:2017zqi} for
models with $N=1,2,3$ to general $N$ and $N_f$ at least up to three-loop order.

\begin{figure*}[t]
	\centering
	\includegraphics{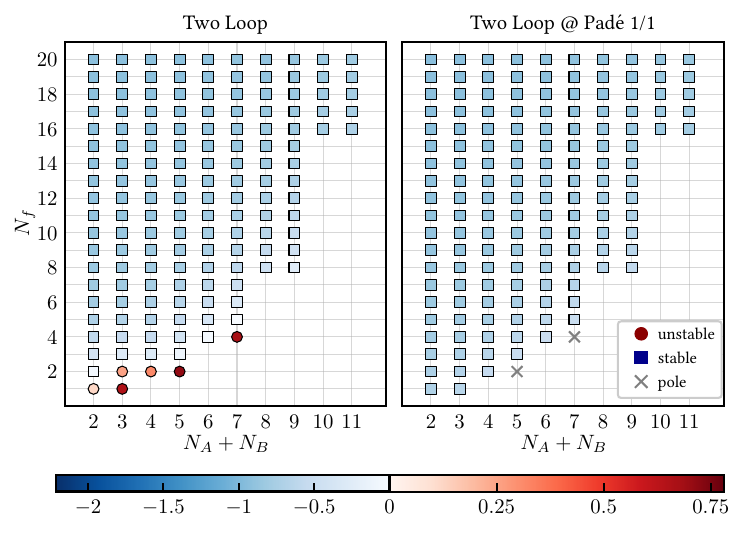}
	\caption{
    Stability analysis of the IFP in $d=3$, i.e., in the limit $\epsilon \to
    1$, at the two-loop order for $N_A + N_B \leqslant 11$ and various values
    of $N_f = d_\gamma/4$ satisfying Eq.~\eq{compatible}.  The stability is
    determined using the na\"ively $\epsilon$-expanded eigenvalues of the
    stability matrix in Eq.~\eqref{eq:stabmat} as well as the Padé
    approximants.  A square marks a stable FP while a circle an unstable FP.
    The color gradient indicates the value of the most marginal eigenvalue,
    i.e., the eigenvalue closest to zero.
    When at least one of $N_A$ or $N_B$ is even there are multiple ways to implement compatibility.
    The models with the minimal value of $N_f$ are only possible via Yukawas of ``Type II'', 
    cf., discussion at the beginning of Sec.~\ref{subsec:compatible}.
    We do not provide a
    prediction if the given Padé approximant exhibits a pole in the interpolation interval
    $\epsilon \in [0,1]$ and mark the respective points with a gray cross. 
	\label{fig:IFP_2L}
}
\end{figure*}

\begin{figure*}[t]
	\centering
	\includegraphics{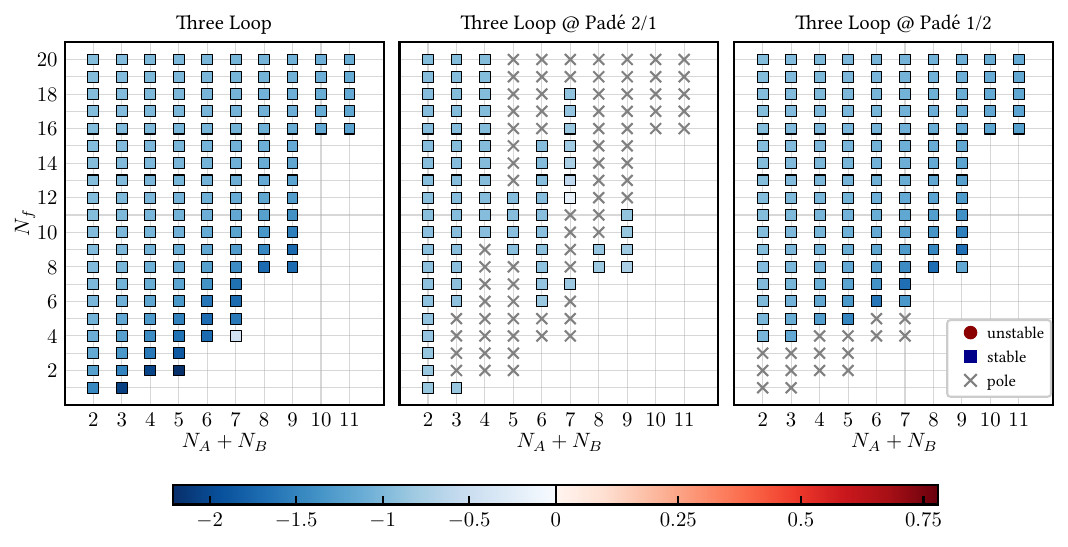}
	\caption{
    Stability analysis of the IFP in $d=3$, i.e., in the limit $\epsilon \to 1$, at
    the three-loop order for $N_A + N_B \leqslant 11$ and various values of
    $N_f = d_\gamma/4$ satisfying Eq.~\eq{compatible}. 
    The stability is determined using the na\"ively $\epsilon$-expanded
    eigenvalues of the stability matrix in Eq.~\eqref{eq:stabmat} as well as the Padé approximants.
    A square marks a stable FP while a circle an unstable FP. 
    The color gradient indicates the
    value of the most marginal eigenvalue, i.e., the eigenvalue closest to zero.
    When at least one of $N_A$ or $N_B$ is even there are multiple ways to implement compatibility.
    The models with the minimal value of $N_f$ are only possible via Yukawas of ``Type II'', 
    cf., discussion at the beginning of Sec.~\ref{subsec:compatible}.
    We do not provide a
    prediction if the given Padé approximant exhibits a pole in the interpolation interval
    $\epsilon \in [0,1]$ and mark the respective points with a gray cross.
	\label{fig:IFP_3L}
  }
\end{figure*}

Employing the $\epsilon$-expansion framework~\cite{WILSON197475}, the values of
the couplings at the IFP admit a perturbative expansion for small $\epsilon$:
\begin{equation}\label{eq:epsExpFP}
	c_m^* = C_m^{(1)} \epsilon + C_m^{(2)} \epsilon^2 + C_m^{(3)} \epsilon^3 + \mathcal{O}(\epsilon^4) \,,
\end{equation}
with $C_m^{(n)}$ determined by the roots, i.e., the fixed-point solutions, of
the $n$-loop $\beta$ functions $\beta_{c_m}\!\left( \{ c^{*}_k \} \right) = 0$.
The full, symbolic expressions for the $C_m^{(n)}$ up to the three-loop order
are given in the attached \texttt{Mathematica} file and in the ancillary files
of the arXiv submission of this paper~\cite{Uetrecht:2025llz}.

\subsubsection{Stability of the isotropic fixed point in \texorpdfstring{$d=3$}{d=3}}
We now consider the $\epsilon$-expansion predictions on the stability of 
the IFP in $d=3$, i.e., by taking the limit $\epsilon\to1$.
The stability is determined from the stability matrix in Eq.~\eq{stabmat}
using the $\epsilon$-expansion predictions of the $\beta$ functions of all five couplings.

For leading values of $N_{A,B,f}$, the stable and unstable IFPs
are summarized in Figs.~\ref{fig:IFP_2L} and \ref{fig:IFP_3L} at
two-loop, ${\cal O}(\epsilon^2)$, and three-loop ${\cal O}(\epsilon^3)$ level, respectively. 
The combinations of $N_f$ and $N$ for which the IFP is stable are represented by
squares and the unstable ones by circles. The color indicates the value of the
eigenvalue $\theta_1$ closest to marginality at $\epsilon = 1$, quantifying how
stable (blue, negative) or unstable (red, positive) each point is. We also
calculate the 2/1 and 1/2 Pad\'e approximants
at the three-loop order, as
well as the 1/1 Pad\'e approximants at the two-loop order. 
In the Pad\'e approximants  we exclude points that exhibit 
poles within the interpolation region $0 \leqslant \epsilon \leqslant 1$.

Before discussing our analysis of the stability of the IFP at higher order in
$\epsilon$, we recall that at leading order, i.e., $\epsilon^1$, the IFP appears 
to be fully stable for all choices of $N_f$ and $N$ that are allowed within the 
constraints of the compatible models~\cite{PhysRevB.97.041117}.
This observation from Ref.~\cite{PhysRevB.97.041117} is in sharp contrast to the
established behavior of purely statistical models with two coupled
$\SO(N_{A,B})$ order parameters and no fermions, where the only choice of $N_A$, $N_B$ leading
to a stable IFP is $N_A=N_B=1$~\cite{PhysRevB.67.054505}.\footnote{
Note that the stability of the IFP at the $N_A = N_B = 1$ only becomes apparent 
once higher orders in $\epsilon$ are taken into account.}
However, the stability of the IFP in the system coupled to Dirac fermions has
not been confirmed for higher-loop orders, yet. In the following, we fill this
gap.

Expanding up to next-to-leading order $\propto\epsilon^2$, we observe that a
direct evaluation of $\epsilon=1$, yields that the IFP is stable for almost all
combinations $(N_f,N)$, see the left panel of Fig.~\ref{fig:IFP_2L}.
In particular, the stability of the IFP can be brought under control by going
to larger values of $N_f$, which aligns with the previous finding that fermions
have a stabilizing effect on the IFP~\cite{PhysRevB.97.041117}.
We study this effect more in Sec.~\ref{sec:largeNf}.
However, for small values of $N_f$, i.e., $N_f = 1,2,4$ we find positive
values of the stability exponent for some values of $N$ upon direct
substitution of $\epsilon=1$, suggesting an unstable IFP.
Starting from $N_f=5$, the IFP is consistently stable for all allowed values of
$N$ at all available orders and for all available Pad\'e approximants.

At order next-to-next-to-leading order $\propto\epsilon^3$, all IFPs are
predicted to be stable in a direct evaluation of $\epsilon=1$, even for smaller
values of $N_f$.  However, here, the $\epsilon^3$ terms provide comparatively
large contributions to the stability exponent, which suggests that the
asymptotic series already behaves badly at this order, see also the next
section.
Therefore, we also calculate the Pad\'e approximants 2/1 and 1/2 to obtain
estimates for the stability exponent. We exclude points that show a
pole within the interpolation interval $\epsilon\in [0,1]$.  The 2/1 Pad\'e approximant
appears to confirm the overall tendency towards a stable IFP and the stability
exponents $\theta_1$ are now much closer to marginality than for the direct evaluation at $\epsilon=1$.
We note that for $N > 3$ there is an increasing number of points invalidated
due to poles in the range $0 \leqslant \epsilon \leqslant 1$.  For 1/2 Padé
all IFPs are predicted as clearly stable, i.e., safely away from marginality. At
small values of $N_f$ and $N$, the predictions are, however, plagued by
singularities, which disappear as $N_f$ increases.

We now specifically discuss the case of $N_f = 2$ due to its relevance for
spin-$\tfrac12$ fermions on the honeycomb lattice of graphene near a quantum
(multi)critical point.
A direct evaluation of the stability exponent at order $\epsilon^2$ with $\epsilon=1$ 
suggests that the IFP is (at least slightly) unstable for $N=3,4,5$.
However, the 1/1 Pad\'e approximant indicates stability for $N=3,4$.
For $N=5$, there is a singularity in the interval $\epsilon\in[0,1]$ and we
therefore do not give an estimate.
Extending the expansion to order $\propto\epsilon^3$, each of the points is decisively stable in a direct evaluation of $\epsilon=1$. 
Using 2/1 Pad\'e approximants, stability prevails for $N=2$, but no 
predictions can be made for $N > 2$ due to poles.
Similarly, the 1/2 Pad\'e approximants, exhibit poles that are present for the entire range in $N$.
In App.~\ref{app:B}, we show, as an example, the five eigenvalues of the compatible $\SO(3) \times \SO(2)$--symmetric model ($N=5$) as a function of $\epsilon$ at two- and three-loop order.
Our interpretation of this analysis is that for $N_f=2$ there is a tendency towards
stability of the IFP also at higher loop orders, which can be inferred from the
corresponding Pad\'e approximants wherever they do not show any poles in
$\epsilon\in [0,1]$.
However, owing to the abundance of poles the results are unfortunately not fully conclusive.

\subsubsection{Chiral \texorpdfstring{$\grO(4)$}{O(4)} and \texorpdfstring{$\grO(5)$}{O(5)} models at \texorpdfstring{$N_f=2$}{Nf=2}}

Critical exponents of the chiral Ising, XY, and Heisenberg models have been
discussed in the past years employing various many-body approaches, e.g.,
perturbative and functional RG, quantum Monte Carlo, and the conformal
bootstrap.  Reasonable agreement on the estimates for the exponents has been
achieved in some cases, e.g., the chiral Ising model at $N_f=2$, while in other
cases there are large discrepancies, e.g., for the chiral Heisenberg model at
$N_f=2$.  Within the $\epsilon$~expansion, it has been observed that the behavior of
the asymptotic series tentatively gets worse when $N$ is increased at fixed
$N_f$, i.e., the coefficients already become large at orders $\epsilon^3$ and
$\epsilon^4$~\cite{Zerf:2017zqi}. 
Before this work the systematic analysis for $N=4,5$ beyond one loop was missing.

While the critical exponents $\eta_\Phi$ and $\eta_\Psi$ directly correspond to
the value of their respective anomalous field dimension on the IFP, the inverse
critical scaling exponent, $1/\nu$, is determined by the relation 
\begin{equation}
    1/\nu = 2 - \gamma_{r}^* \,,
\end{equation}
with $\gamma_{r}^* = \gamma_{r_{A,B}}^*$ the value of the anomalous square boson mass 
dimension evaluated on the IFP.
Lastly, we also determine the leading corrections-to-scaling exponent $\omega_1 = - \theta_k$, with $\theta_k$ being the relevant eigenvalue closest to marginality.

\begin{widetext}
For the chiral $\grO(4)$ model at $N_f = 2$ and its leading critical exponents, we find the series expansion
\begin{equation}\label{eq:critical-O(4)}    
\begin{aligned}
    1/\nu^{\grO(4)} &= 2 - 2\epsilon + \frac{67}{120} \epsilon^2 + \frac{16631-44600\zeta_3}{24000} \epsilon^3 
        \approx 2 - 2.0 \epsilon + 0.5583 \epsilon^2 - 1.541 \epsilon^3 \,, \\
    \nu^{\grO(4)} &= \frac{1}{2} + \frac{1}{2} \epsilon + \frac{173}{480} \epsilon^2 + \frac{4569+44600\zeta_3}{96000} \epsilon^3 
        \approx 0.5 + 0.5 \epsilon + 0.3604 \epsilon^2 + 0.6060 \epsilon^3 \,, \\
    \eta^{\grO(4)}_{\Phi} &= \epsilon + \frac{3}{8} \epsilon^2 + \frac{23}{240} \epsilon^3 
        \approx 1.0 \epsilon + 0.375 \epsilon^2 + 0.0958 \epsilon^3 \,, \\
    \eta^{\grO(4)}_{\Psi} &= \frac{1}{2} \epsilon - \frac{1}{24} \epsilon^2 - \frac{943}{2880} \epsilon^3 
        \approx  0.5 \epsilon - 0.0416 \epsilon^2 - 0.3274 \epsilon^3 \,, \\
    \omega^{\grO(4)}_{1} &= \epsilon + \frac{19}{24}\epsilon^2 + \frac{1993 + 540 \zeta_3}{1440} \epsilon^3
        \approx 1.0\epsilon - 0.792 \epsilon^2 + 1.835 \epsilon^3  \,.
\end{aligned}
\end{equation}
For the chiral $\grO(5)$ model at $N_f = 2$, we find
\begin{equation}\label{eq:critical-O(5)} 
\begin{aligned}
   1/\nu^{\grO(5)}  &=2 - \frac{36}{13}\epsilon - \frac{1297}{4394} \epsilon^2 - \frac{76299607 + 76556688 \zeta_3}{40099644} \epsilon^3 
        \approx 2 - 2.7692  \epsilon - 0.2952 \epsilon^2 - 4.1977 \epsilon^3 \,, \\
    \nu^{\grO(5)} &=\frac{1}{2} + \frac{9}{13} \epsilon + \frac{18145}{17576} \epsilon^2 + \frac{321968719 + 76556688\zeta_3}{160398576} \epsilon^3
        \approx 0.5 + 0.6923 \epsilon + 1.0324 \epsilon^2 + 2.5810 \epsilon^3 \,, \\
    \eta^{\grO(5)}_\Phi &=\frac{4}{3} \epsilon + \frac{1451}{1014} \epsilon^2 + \frac{24771515 - 10967424 \zeta_3}{6169176} \epsilon^3 
        \approx 1.3333 \epsilon + 1.4309 \epsilon^2 + 1.8784 \epsilon^3 \,, \\
    \eta^{\grO(5)}_\Psi &=\frac{5}{6} \epsilon + \frac{1705}{4056} \epsilon^2 + 5 \frac{2201825 - 2741856 \zeta_3}{12338352} \epsilon^3 
        \approx 0.8333 \epsilon + 0.4204 \epsilon^2 - 0.4434 \epsilon^3 \,, \\
    \omega^{\grO(5)}_{1} &= \epsilon - \frac{6947}{4056}\epsilon^2 - \frac{330825 - 1942148 \zeta_3}{685464} \epsilon^3
        \approx 1.0\epsilon - 1.7128 \epsilon^2 + 2.9232 \epsilon^3 \,,
\end{aligned}
\end{equation}
where $\eta_\Phi = 2 \gamma_\Phi^*$ and $\eta_\Psi = 2 \gamma_\Psi^*$, cf., App.~\ref{app:anomDimComp}.
\end{widetext}

The magnitude of the coefficients $\propto \epsilon^3$ of almost all critical
exponents in Eqs.~\eq{critical-O(4)} and \eq{critical-O(5)} 
increases with respect to the previous order $\propto \epsilon^2$, pointing towards a radius of convergence $\epsilon < 1$. 
Thus, the validity of the limit $\epsilon \to 1$ is called into question.
This is in agreement with the larger values of the most marginal eigenvalue for these models,
as indicated in Figs.~\ref{fig:IFP_2L} and \ref{fig:IFP_3L}.  We therefore
conclude that --- at least for small values of $N_f$, e.g., $N_f=2$ --- the
series expansion in $\epsilon$ results in a badly behaved asymptotic series
already at low expansion orders. Consequently, we refrain from providing
resummed estimates for the exponents in $d=3$. 
A better understanding of the
quantitative behavior of the asymptotic series may facilitate the extraction of
estimates in the future. 

Alternatively, the behavior of the asymptotic series can be brought under
control by considering larger values of $N_f$ as we will discuss in the next
section.

\subsubsection{Large \texorpdfstring{$N_f$}{Nf}\label{sec:largeNf}}

In this section, we examine the compatible model in a large-$N_f$ limit. In
doing so, we can develop an analytical understanding for many of the fixed points
in Figs.~\ref{fig:IFP_2L} and~\ref{fig:IFP_3L} including their stability.
For one, this limit describes the compatible model at any value $N$, but for a
larger number of fermion flavors $N_\Psi \gg N$.  Secondly, the large-$N_f$
limit also characterizes the case when $N$ is large, even if the number of
fermions is minimal as their number grows fast, i.e., $N_{f}^{\text{min}}
\propto 2^N \gg N$. 
Thus, the analysis here predicts the FP stability of both the top row and the
rightmost points of Figs.~\ref{fig:IFP_2L}~and.~\ref{fig:IFP_3L}.

To facilitate the limit, we introduce versions of the couplings $g$ and
$\lambda$ rescaled by powers of $N_f$ such that their respective $\beta$
functions neither diverge nor vanish completely as $N_f \to \infty$. Instead,
the leading large-$N_f$ contributions are projected out, while others are
suppressed by powers of $1/N_f$.
Each loop order can contribute a factor $\lambda$ or $g^2$ and up to one
fermion bubble, which yields a factor $N_f$. Thus, the rescaled couplings read 
\begin{equation}
    G = \frac{N_f g^2}{(4\pi)^2} \qquad \text{ and } \qquad \Lambda = \frac{N_f \lambda}{(4\pi)^2} \,.
\end{equation}
Their leading-order $\beta$ functions in the large-$N_f$ expansion are found
exactly from the one-loop results in Eqs.~\eq{betag2Compatible} and
\eq{betalamCompatible}
\begin{align}\label{eq:largeNfResc}
    \beta_{G} =-\epsilon G + 4 G^2 \,,\quad \beta_{\Lambda} = -\epsilon \Lambda - 16 G^2 + 8 G \Lambda \,,
\end{align}
while all higher loops only contribute to subleading corrections $\propto 1/N_f$.
Thus, we obtain a non-trivial fixed point solution
\begin{equation}\label{eq:largeNfFP}
    G^{*} = \epsilon/4  + \mathcal{O}\left(\frac1{N_f}\right)\,,\qquad\Lambda^* = 2 \epsilon + \mathcal{O}\left(\frac1{N_f}\right)\,,
\end{equation}
which is now an expansion in $1/N_f$ rather than $\epsilon$.
Evaluating the corresponding full stability matrix of all five marginal
couplings (after rescaling them like Eq.~\eqref{eq:largeNfResc}), at the fixed
point $(G^{*},\Lambda^*)$, we find that all eigenvalues are equal
$\theta_{1,2,3,4,5} = -\epsilon$. 
Thus, the large-$N_f$ fixed point is physical and stable for all $\epsilon > 0$.
An analogous computation of the critical exponents of the fixed point in the
large-$N_f$ limit produces
\begin{equation}\label{eq:critLargeNf}
\begin{aligned}
    1/\nu^{(N_f\to\infty)} &= 2 - \epsilon \,, \\
    \eta^{(N_f\to\infty)}_\Phi &= \epsilon \,, \\\
    \eta^{(N_f\to\infty)}_\Psi &= 0 \,, \\
    \omega_1^{(N_f\to\infty)} &= \epsilon\,.
\end{aligned}    
\end{equation}
Already at $N=10$ and $N_{f,\text{min}} = 16$, the prediction of the
critical exponents via the large-$N_f$ as well as the $\epsilon$-expansion are in good agreement.

In summary, for the compatible model our results indicate that a stable IFP may
exist for all allowed values of $N$ and $N_f$. 
As $N_f$ grows exponentially with increasing $N$, the stability of the IFP is
corroborated by the large-$N_f$ fixed point~\eq{largeNfFP}, which is stable for
all values of $\epsilon$. This claim is supported further by the predictions of
all pole-free Pad\'e resummations at the three-loop order.\\

\subsection{Incompatible \texorpdfstring{$\SO(3) \times \SO(3)$}{SO(3) x SO(3) } Model\label{sec:incompatible}}

Next, we analyze a special case of an incompatible version of Eq.~\eq{bareLagr4d} 
in its symmetric subspace, i.e.,
\begin{equation}
\begin{aligned}
  \mathrm{Re}(y_A)= \mathrm{Re}(y_B) &= g\,,\quad
  \mathrm{Im}(y_A)= \mathrm{Im}(y_B) = 0\,,\\
  \lambda_A       = \lambda_B &= \lambda\,,\quad
  \lambda_{AB}    =\lambda_c\,.
\end{aligned}
\end{equation}
In particular, we study the $\SO(3) \times \SO(3)$ symmetric field theory
corresponding to $N_A = N_B = 3$ and $d_\gamma/4 = N_f= 2$.
In the context of interacting electrons on the honeycomb lattice, the model
describes the low-energy Dirac excitations of the semimetallic state coupled to
two triplets of order parameters , i.e., a Néel state and a combined
superconducting-CDW state~\cite{PhysRevLett.128.117202}.
A recent large-scale QMC analysis of this system~\cite{PhysRevLett.128.117202}
found evidence for a scaling collapse, suggesting the presence of a stable RG
fixed point in the corresponding field theory. 
RG analyzes at one-~\cite{Herbut:2022zzw} and two-loop order~\cite{Uetrecht:2023uou},\footnote{
In Ref.~\cite{Uetrecht:2023uou}, we normalized the Gamma matrices appearing
in the Yukawas with an additional factor of $1/2$, i.e., the definition of $g_{A,B}$ differs by a factor $1/2$.
}
instead, could not corroborate this observation.
Here, we extend the previous analyses to three loops.

Retaining $N_f$ as a general parameter, the $\beta$ functions of the symmetric
subspace at three-loop order read
\begin{widetext}
\begin{align}\label{eq:incompatible-beta-1}
    	\beta_{g^2} =-\epsilon g^2
		  & + \frac{1}{16 \pi^2} 4 g^4(5+N_f) \notag                                                                                                                                                                                            \\
		  & + \frac{1}{(16 \pi^2)^2} \frac{g^2}{2} \left[ -4 g^4 \left( 48 N_f + 49 \right) - 8 g^2 \left( 5 \lambda + 9 \lambda_c \right) + 5 \lambda^2 +3 \lambda_c^2 \right] \nonumber                                        \\
		  & + \frac{1}{(16 \pi^2)^3} \frac{g^2}{8} \Big[ 48 (5 (26 + 5 N_f) g^4 \lambda +  3 (46 + 15 N_f) \lambda_c) + 2 g^2 ((5 - 75 N_f) \lambda^2 +600 \lambda \lambda_c + 3 (41 - 15 N_f) \lambda_c^2)\nonumber                                                                              \\
		  & \qquad - 55 \lambda^3 - 45 \lambda \lambda_c^2 - 12 \lambda_c^3 + 8 \left(-983 + 979 N_f + 160 N_f^2 + 144 (9 + 5 N_f) \zeta_3\right) g^6  \Big]\,,
\end{align}
\begin{align}
\beta_{\lambda}= -\epsilon\lambda
          & + \frac{1}{16 \pi^2} \left[ -16 N_f g^4+ 8 N_f g^2 \lambda +11 \lambda ^2+3 \lambda_c^2 \right] \notag    \\                                                    
          & + \frac{1}{(16 \pi^2)^2} \left[ 512 N_f g^6 + 24 N_f g^4 \left(2 \lambda_c - 3\lambda\right) - 4 N_f g^2 \left(11 \lambda^2+3 \lambda_c^2\right) - 3 \left(23 \lambda^3 + 5 \lambda \lambda_c^2+4 \lambda_c^3\right) \right] \nonumber \\ 
		  & + \frac{1}{(16 \pi^2)^3} \frac{1}{32} \Big[ 16 N_f g^2 (517 \lambda^3 + 99 \lambda \lambda_c^2 + 96 \lambda_c^3) + 64 N_f g^6 \big( 48 \lambda_c (-51 + 8 N_f - 48 \zeta_3) + \lambda (-4083 + 740 N_f - 48 \zeta_3) \big) \nonumber  \\
		  & \qquad - 256 N_f (499 + 622 N_f + 384 \zeta_3) g^8 + 4 \big( 294 \lambda^2 \lambda_c^2 + 24 \lambda \lambda_c^3 (137 + 48 \zeta_3) + 3 \lambda_c^4 (25 + 96 \zeta_3) + \lambda^4 (6023 + 3552 \zeta_3) \big) \nonumber                \\
		  & \qquad + 32 N_f g^4 \big( 3 \lambda_c^2 (241 - 12 N_f + 180 \zeta_3) + \lambda^2 (2111 - 132 N_f + 636 \zeta_3) + 12 \lambda (\lambda_c + 102 \lambda_c \zeta_3) \big) \Big] \,,
\end{align}
\begin{align}\label{eq:incompatible-beta-2}
		\beta_{\lambda_c}=-\epsilon  \lambda _c
		  & + \frac{1}{16 \pi^2} \left[-48 N_f g^4 + 8 N_f g^2 \lambda_c +2 \lambda_c (5 \lambda +2 \lambda_c) \right]\notag                                                                                                                   \\
		  & +\frac{1}{(16 \pi^2)^2} \left[ 1024 N_f g^6 + 40 N_f g^4 \left(2 \lambda -\lambda_c\right)-8 N_f g^2 \lambda_c (5 \lambda +2 \lambda_c)-\lambda_c (5 \lambda +\lambda_c) (5 \lambda +11 \lambda_c) \right]                  \nonumber \\
		  & + \frac{1}{(16 \pi^2)^3} \frac{1}{32} \Big[ 16 N_f g^2 \lambda_c (165 \lambda^2 + 480 \lambda \lambda_c + 67 \lambda_c^2) \nonumber                                                                                                \\
		  & \qquad+ 64 N_f g^6 \big( \lambda_c (-6739 + 996 N_f - 1584 \zeta_3) + 80 \lambda (-51 + 8 N_f - 48 \zeta_3) \big) - 256 N_f (25 + 1290 N_f + 768 \zeta_3) g^8 \nonumber                                                                  \\
		  & \qquad+ 16 \lambda_c \big( 780 \lambda^3 + 30 \lambda \lambda_c^2 (27 + 16 \zeta_3) + 3 \lambda_c^3 (55 + 16 \zeta_3) + 5 \lambda^2 \lambda_c (133 + 144 \zeta_3) \big) \nonumber                                                 \\
		  & \qquad+ 64 N_f g^4 \big( 10 \lambda^2 (53 + 15 \zeta_3) + 15 \lambda \lambda_c (57 - 4 N_f + 76 \zeta_3) + 2 \lambda_c^2 (347 - 12 N_f + 171 \zeta_3) \big) \Big] \,.                                                  
\end{align}
\end{widetext}

In~Ref.~\cite{Herbut:2022zzw}, we found that at one-loop order the above
Eqs.~\eq{incompatible-beta-1}--\eq{incompatible-beta-2} only have fixed-point
solutions for $N_f<N_c^<\approx 0.0164$ or for $N_f>N_c^>\approx 16.83$, but
not for the case of $N_f$, which was studied in QMC
simulations~\cite{PhysRevLett.128.117202}.
The lower and the upper critical flavor numbers, i.e., $N_c^<$ and $N_c^>$,
respectively, will be subject to corrections at higher loop orders.
The magnitude of these corrections is controlled by the deviation of the
spacetime dimension from four, i.e., by the value of $\epsilon$ in
$d=4-\epsilon$.

To determine these corrections systematically, we follow the same strategy we
already put forward in
Refs.~\cite{PhysRevB.100.134507,PhysRevLett.78.980,Uetrecht:2023uou}: {\itshape
i)}~We first numerically calculate  solutions of the three-loop $\beta$
functions in the $\epsilon-N_f$~plane, i.e., we vary $\epsilon$ and $N_f$ as
continuous parameters.
{\itshape ii)}~We then follow the boundaries separating the regions where fixed
points exist from regions without fixed points.  For small $\epsilon$ the
boundary values for $N_c^{\lessgtr}$ converge to their one-loop estimate.
We then determine the corrections up to order $\mathcal{O}(\epsilon^2)$,
corresponding to the three-loop equations, by extrapolating the boundary line
from close to $\epsilon=0$.

\begin{figure}[t]
    \centering
    \includegraphics{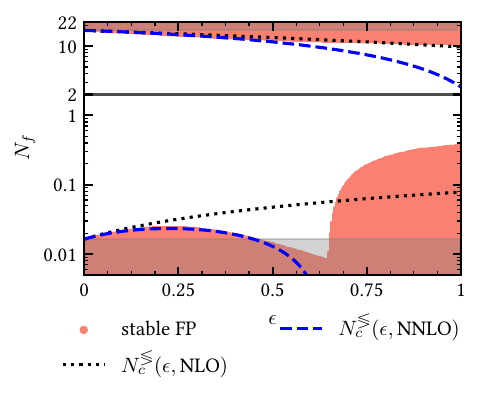}
    \caption{Red dots mark values of $\epsilon$ and $N_f$ where a stable, physical fixed-point exists at three-loop order. For comparison, the gray shaded areas show the regions, where a FP solution already exists within the one-loop $\beta$ functions. Extrapolating the boundaries from small $\epsilon$, we obtain numerical estimates for $N_{c}^{\lessgtr}$, cf., Eqs.~\eqref{eq:Nf_c_gt} and~\eqref{eq:Nf_c_lt}, including the three-loop (NNLO) corrections, yielding the dashed, blue lines. For comparison, the two-loop (NLO) results~\cite{Uetrecht:2023uou} are displayed as dotted, black lines.}
     \label{fig:symNf}
\end{figure}

Our results are summarized in Fig.~\ref{fig:symNf}, where a red dot marks a
stable fixed-point solution, and we excluded unphysical fixed points with $g^{*
2} < 0$ or $\lambda^* < 0$. 
For comparison, we include the areas of stable physical fixed points as
predicted by the one-loop order~\cite{Herbut:2022zzw} in gray.  For the
extrapolation procedure, we resort to our two-loop
analysis~\cite{Uetrecht:2023uou}, which suggests the ans\"atze
\begin{align}\label{eq:Nf_c_gt}
    N_{c}^{>} &\approx 16.83 - 7.14 \epsilon + \delta N_{c}^{>\,(\text{3-loop})} \epsilon^2 +\mathcal{O}(\epsilon^3)\,,\\[1em]
\label{eq:Nf_c_lt}
    N_{c}^{<} &\approx 0.0164 + 0.062 \epsilon + \delta N_{c}^{< \,(\text{3-loop})} \epsilon^2 +\mathcal{O}(\epsilon^3)\,,
\end{align}
where the numerical values of the $\mathcal{O}(\epsilon^1)$~coefficients are
taken from Ref.~\cite{Uetrecht:2023uou}, which we have consistently rederived
here. The $\delta N_{c}^{\lessgtr\,(\text{3-loop})}$ need to be determined by a
fit to the numerical data of Fig.~\ref{fig:symNf} for small values of
$\epsilon$.

For this fit, we consider values up to $\epsilon^{(\text{3-loop})}_{\mathrm{max}} =
0.4$.
Numerically searching for fixed-point solutions starting at
$\epsilon_{\mathrm{start}} = 0.004$ with step size $\Delta \epsilon = 0.004$,
along with evaluating the three-loop $\beta$ functions for 200 evenly-spaced
logarithmic values in the ranges $1/2 \leq N_f \leq 22$ for $\delta
N_{c}^{>\,(\text{3-loop})}$, and $0.005 \leq N_f \leq 0.5$ for $\delta N_{c}^{<
\,(\text{3-loop})}$, yields
\begin{equation}
     \delta N_{c}^{>\,(\text{3-loop})} \approx -7.12 \,,\quad \delta N_{c}^{<\,(\text{3-loop})} \approx -0.1380 \,.
\end{equation}
The resulting Eqs.~\eq{Nf_c_gt} and \eq{Nf_c_lt} are depicted In
Fig.~\ref{fig:symNf} as the dashed, blue lines. 
In Tab.~\ref{tab:3dFPGraphene}, we have summarized the predictions of the
critical number of fermions for the physically relevant case of $\epsilon \to
1$ and $N_f = 2$ at the one-, two- and three-loop order.

In summary, the three-loop corrections further suppress the critical value
$N_{c}^{>}$ towards $N_f = 2$ in the $d=3$ case.  Therefore, it seems possible
that the true critical flavor number, $N_{f,c}$, is close to or even below $N_f = 2$. 
This would render our results compatible with the numerical observation of
critical (if $N_{c}^{>}<2$) or at least with pseudo-critical walking-like
behavior (if $N_{c}^{>}\gtrsim 2$), cf.,
Refs.~\cite{PhysRevD.80.125005,Gorbenko:2018ncu,Hawashin:2023fnu} and the
related discussion in Ref.~\cite{Uetrecht:2023uou}.

\begin{table}[t]
\begin{tabular*}{\linewidth}{@{\extracolsep{\fill} } ccrr}
  \hline\hline\\[-0.75em]
Loop order & Ref. & $N_c^{>}$  & $N_c^{<}$\\ \hline\\[-0.75em]
$1$ &\cite{Herbut:2022zzw}   & $16.83$  & $0.0164$\\
$2$ &\cite{Uetrecht:2023uou} & $9.69$  & $0.0784$\\
$3$ &this work               & $2.57$  & $-0.0596$\\
\hline
\end{tabular*}
\caption{ Predictions for the critical numbers of fermions,
$N_{c}^{\lessgtr}$, above, or below respectively, which a stable
fixed-point solution exists in $d=3$. 
At $L$-loop order, this is achieved by evaluating 
Eqs.~\eqref{eq:Nf_c_gt} and~\eqref{eq:Nf_c_lt} up to $\epsilon^{L-1}$, and taking the limit $\epsilon \to 1$. 
Utilizing the $1/1$ Padé
approximant for the three-loop results produces identical
predictions.\label{tab:3dFPGraphene}}
\end{table}
%

\section{Conclusions\label{sec:conclusion}}

In this work, we have constructed a $(4-\epsilon)$--dimensional, $\SO(N_A)
\times \SO(N_B) \times \mathrm{U}(N_\Psi)$--symmetric quantum field theory
that serves as a dimensional continuation for several $(2+1)$-dimensional
Gross--Neveu--Yukawa (GNY) theories with two order parameters in different
symmetrical subspaces. 
We have computed three-loop $\beta$ functions and anomalous dimensions,
providing unprecedented access to fixed-point values, eigenvalues of the
stability matrix, and critical exponents up to order $\sim \epsilon^3$.

In particular, we have investigated the occurrence and stability of isotropic
fixed points (IFPs) in this class of GNY models that are compatible with the
emergence of an $\SO(N_A\!+\! N_B)$ symmetry.
The $\epsilon$-expansion and Pad\'e resummations corroborate the stability of
the symmetry-enhanced IFPs for large~$N_f$. 
For lower $N_f$, including the physically relevant cases $N_f=2,4$, we observe a
clear tendency towards stability of the symmetry-enhanced IFP wherever we can
extrapolate the $4-\epsilon$ results to the case of 2+1 dimensions using
pole-free Pad\'e approximants.
Nevertheless, the presence of poles in some Pad\'e approximants for
small $N_f$ prevents a fully satisfactory conclusion on this matter.

Considering the case $N_f=2$, we have extracted the series expansion of
the leading critical exponents for the symmetry enhanced chiral $\SO(4)$ and
$\SO(5)$~models up to third order in~$\epsilon$.  
Notably, we find a tendency towards rapidly growing expansion coefficients
at higher orders, rendering a reliable extrapolation to $\epsilon=1$ difficult.
We note that this unfavorable asymptotic behavior can, again, be mitigated by
choosing larger~$N_f$.

Finally, we have revisited a model with $\SO(4) \simeq \SO(3) \times \SO(3)$
symmetry, which was recently suggested to describe criticality of
antiferromagnetism and superconductivity in Dirac  systems based on QMC
data~\cite{PhysRevLett.128.117202}.
Here, we have extended the two-loop analysis from Ref.~\cite{Uetrecht:2023uou}
to three loops.
While we did not find an admissible RG fixed point facilitating true critical
behavior in the relevant $N_f=2$~version of the model, we found indications
that a fixed-point collision appears for slightly higher $N_f$.
This suggests that the model could show pseudo-critical walking
behavior~\cite{PhysRevD.80.125005,Gorbenko:2018ncu}, which can be extremely
difficult to distinguish from true scaling with large corrections in numerical
calculations.
This could explain the observation of apparent scaling in
Ref.~\cite{PhysRevLett.128.117202}.

In view of the recent experimental observation of a relativistic Mott
transition in a highly-tunable Dirac system based on twisted WSe${}_2$
tetralayers~\cite{ma2024relativistic}, it will be interesting to explore
whether the system can be further engineered to be close to a multicritical
point where two compatible orders meet.
This could be a route for the experimental observation of enhanced symmetries
in Dirac systems at a quantum multicritical point.

Also, in the context of theoretical studies of deconfined quantum critical points as realized in fermionic setups with Dirac excitations~\cite{Liu_2019,PhysRevLett.119.197203}, the multicritical points discussed in our work are relevant for the understanding of the corresponding quantum phase diagrams.

\begin{acknowledgments}
I.F.H.~is supported by the NSERC of Canada. M.M.S., E.S., and M.U. are
supported by the Mercator Research Center Ruhr under Project No.
Ko-2022-0012. M.M.S. acknowledges funding from the Deutsche
Forschungsgemeinschaft (DFG, German Research Foundation) within Project-ID
277146847, SFB 1238 (Project No.~C02), and the DFG Heisenberg programme
(Project-ID 452976698). M.U.~is supported by the doctoral scholarship program
of the \textit{Studienstiftung des deutschen Volkes}, and is grateful to
SISSA for the kind hospitality during the final stages of this work.
\end{acknowledgments}

\appendix

\begin{widetext}
\section{Anomalous Dimensions of the Compatible Model}\label{app:anomDimComp}
In this Appendix, we collect the anomalous dimensions of the scalars,
$\gamma_\Phi = \gamma_{\Phi_{A,B}}$, the fermions, $\gamma_\Psi$, and the
scalar squared mass, $\gamma_r = \gamma_{r_{A,B}}$, of the compatible model, i.e., the
Lagrangian~ in Eq.~\eq{bareLagr4d} constrained to satisfy the compatibility
condition~\eq{betacom} via Eq.~\eq{compatible-4d}, for general
values of $N = N_A + N_B$ and $N_f$. The following results may also be derived
from the attached \texttt{Mathematica} file, which contains the three-loop RGEs
for the full Lagrangian in Eq.~\eq{bareLagr4d}.

\begin{align}
    \gamma_{\Phi} =  \frac{d \log Z^{1/2}_\Phi}{d\log\mu} = & \frac{g^2 N_f}{8 \pi^2} 
    + \frac{1}{(16 \pi^2)^2} \left[-4 N_f g^4 - N N_f g^4 + \frac{\lambda^2}{2} + \frac{N \lambda^2}{4} \right] \notag \\
    & + \frac{1}{(16 \pi^2)^3} \Bigg[ 32 N_f g^6 - 36 N N_f g^6 + \frac{53}{8} N^2 N_f g^6 + 18 N_f^2 g^6 + 7 N N_f^2 g^6 \notag \\
    & \qquad + 10 N_f g^4 \lambda + 5 N N_f g^4 \lambda - \frac{15}{4} N_f g^2 \lambda^2 - \frac{15}{8} N N_f g^2 \lambda^2 - \lambda^3 - \frac{5 N \lambda^3}{8} - \frac{N^2 \lambda^3}{16} \notag \\
    & \qquad - 24 N_f \zeta_3 g^6 + 36 N N_f \zeta_3 g^6 - 6 N^2 N_f \zeta_3 g^6 \Bigg] \,, \label{eq:gammaPhi} \\[8pt]
    \gamma_{\Psi} =  \frac{d \log Z^{1/2}_\Psi}{d\log\mu} = & \frac{g^2 N}{32 \pi^2} 
    - \frac{1}{(16 \pi^2)^2} \left[ \frac{N^2}{8} g^4 + \frac{3N N_f}{2} g^4  \right] \notag \\
    & + \frac{1}{(16 \pi^2)^3} \frac{1}{8} \Bigg[ 32 N (2 + N) g^4 \lambda - 11 N (2 + N) g^2 \lambda^2 \notag \\
    & \qquad + N (192 - 248 N + 41 N^2 + 80 N_f + 108 N N_f - 48 N_f^2 - 48 (4 + (-6 + N) N) \zeta_3) g^6 \Bigg] \,, \label{eq:gammaFermion} \\[8pt]
    \gamma_{r} =  \frac{d \log r}{d\log\mu} = & \frac{1}{16 \pi^2} \left[ 4 N_f g^2 + (2 + N) \lambda \right]  
    - \frac{1}{(16 \pi^2)^2} \left[ 4 (-4 + 5 N) N_f g^4 + 8 (2 + N) N_f g^2 \lambda + 5 (2 + N) \lambda^2 \right] \notag \\
    & + \frac{1}{(16 \pi^2)^3} \Bigg[ 66 (2 + N) N_f  g^2 \lambda^2 + 6 (2 + N) (37 + 5 N) \lambda^3 \notag \\
    & \qquad + 4 (2 + N) N_f (-8 + 39 N - 24 N_f - 24 (-6 + N) \zeta_3) g^4 \lambda \notag \\
    & \qquad + 2 N_f (400 N_f - 192 (4 + \zeta_3) + N (-352 - 11 N + 312 N_f - 48 (-14 + N) \zeta_3)) g^6 \Bigg] \,. \label{eq:gammaScalarMass}
\end{align}
In the above, $Z_\Psi$ and $Z_\Phi$ denote renormalization constants, which are
related to the bare fields as $\Psi^{(0)} = Z_\Psi^{1/2} \Psi$ and
$\vec{\Phi}^{(0)} = Z_{\Phi}^{1/2} \vec{\Phi}$.

\section{Example of a compatible \texorpdfstring{$\SO(3) \times \SO(2)$}{SO(3) x SO(2)} model for \texorpdfstring{$N_f = 2$}{Nf = 2}}\label{app:B}

In Fig.~\ref{fig:SO5_EVs} of this Appendix, we show the five eigenvalues of the stability matrix in Eq.~\eqref{eq:stabmat} for the compatible $\SO(3) \times \SO(2)$--symmetric theory in Eq.~\eqref{eq:Lag3d}.
On the honeycomb lattice of graphene, having $N_f = 2$ fermions, the two compatible order parameters correspond, e.g., to a three-component antiferromagnetic Néel order parameter and a two-component Kekulé valence-bond solid order parameter, cf., Eq.~\eqref{eq:AFMVBS}.

\newpage

\begin{figure}
     \centering
     \includegraphics{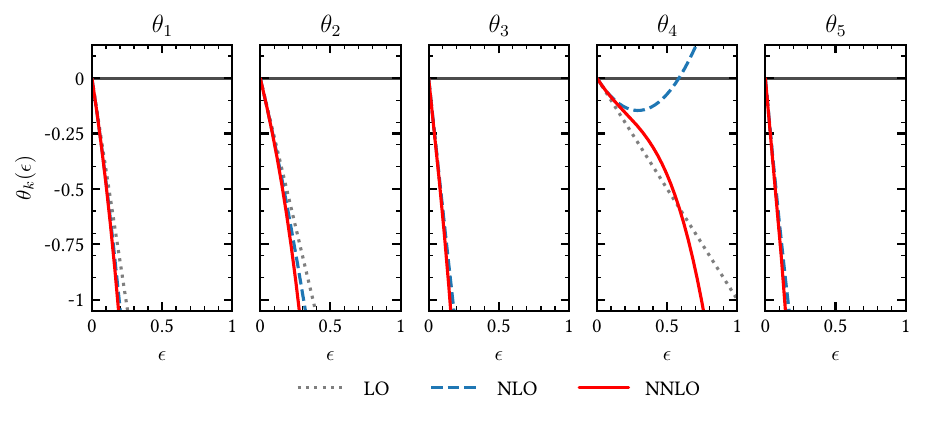}
     \caption{
       The five eigenvalues $\theta_k$ of the stability matrix in Eq.~\eqref{eq:stabmat} as a function of $\epsilon = 4-d$,
       evaluated at the $\SO(5)$--symmetric isotropical fixed point of the compatible $\SO(3) \times \SO(2)$ model.
       The gray dotted, blue dashed, and solid red lines correspond to 
       the one-loop (LO)~\cite{PhysRevB.97.041117}, two-loop (NLO), and three-loop (NNLO) results, respectively.
       Analytically continuing the $\theta_k$ to $d=2+1$, i.e., taking the limit $\epsilon \rightarrow 1$, the IFP is stable (IR attractive) both at LO and NNLO.
       The eigenvalue at NNLO closest to marginality in $d=2+1$ is $\theta_4 \approx -2.21$.
       Accordingly, the stable $N_A + N_B = 5$ IFP in Fig.~\ref{fig:IFP_3L} is indicated as a blue square, where the gradient encodes the value of $\theta_4$. At NLO and in the limit $\epsilon \to 1$, $\theta_4 \approx 0.72 > 0$ making the IFP unstable (IR repulsive).
 \label{fig:SO5_EVs}}
\end{figure}

\end{widetext}

\addcontentsline{toc}{section}{References}
\bibliography{references}
\bibliographystyle{JHEP}

\end{document}